%% file: neurips_2026.tex
\documentclass{article}

\PassOptionsToPackage{numbers,sort&compress}{natbib}

\usepackage[final,main]{neurips_2026}

\usepackage[utf8]{inputenc}
\usepackage[T1]{fontenc}
\usepackage{placeins}
\usepackage{amsmath}
\usepackage{amssymb}
\usepackage{mathtools}
\usepackage{amsthm}
\usepackage{graphicx}
\usepackage{subcaption}
\usepackage{wrapfig}
\usepackage{booktabs}
\usepackage{multirow}
\usepackage{array}
\usepackage{makecell}
\usepackage{colortbl}
\usepackage{xcolor}
\usepackage{caption}
\usepackage{bm}
\usepackage{pifont}
\usepackage{dsfont}
\usepackage{bbm}
\usepackage{fontawesome}
\usepackage{algorithm}
\usepackage{algorithmic}
\usepackage{nicefrac}
\usepackage{microtype}
\usepackage{url}
\usepackage{hyperref}
\usepackage[capitalize,noabbrev]{cleveref}
\usepackage[textsize=tiny]{todonotes}
\usepackage[normalem]{ulem}

\graphicspath{{figures/}}

\definecolor{tprgood}{RGB}{220,240,220}
\definecolor{tprmid}{RGB}{255,245,210}
\definecolor{mypink}{HTML}{FB2E99}
\definecolor{titlelink}{HTML}{2F5B8C}
\definecolor{Orange}{RGB}{245,129,55}
\definecolor{Green}{HTML}{5E9C88}
\definecolor{Blue}{HTML}{4C72B0}

\newcommand{\ours}{\textbf{\textsc{OpenMark}}}
\newcommand{\oursx}{\textbf{\textsc{OpenMarkX}}}
\definecolor{omOpen}{HTML}{231A15}
\definecolor{omMark}{HTML}{8B4B2E}
\newcommand{\oursTitle}{\texorpdfstring{{\raisebox{-.07em}{\includegraphics[height=.87em]{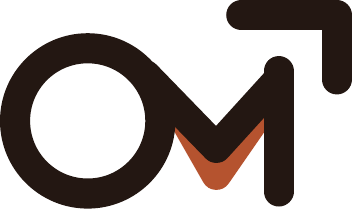}}\kern.18em\fontfamily{Montserrat-TLF}\fontseries{eb}\selectfont\textcolor{omOpen}{Open}\textcolor{omMark}{Mark}\textcolor{omOpen}{X}}}{OpenMarkX}}
\newcommand{\lossnum}[1]{\ding{\numexpr181+#1\relax}}

\newcommand{\improve}[1]{\textcolor{green!50!black}{\small $#1$}}

\DeclareRobustCommand{\revision}[1]{\ifvmode{#1\par}\else{#1}\fi}
\theoremstyle{plain}
\newtheorem{theorem}{Theorem}[section]

\theoremstyle{definition}

\theoremstyle{remark}

\title{\fontsize{16}{19}\selectfont\oursTitle: An Extensible Platform for Watermarking Open-Source Model and Agent Outputs}

\author{
Jiaqi Xue\textsuperscript{\texttt{1}}\thanks{Equal contribution.}, Yifei Zhao\textsuperscript{\texttt{1}}\footnotemark[1], Mansour Al Ghanim\textsuperscript{\texttt{1}}, Shangqian Gao\textsuperscript{\texttt{2}}\\
\textbf{
Ruimin Sun\textsuperscript{\texttt{3}},
Qian Lou\textsuperscript{\texttt{1}},
Mengxin Zheng\textsuperscript{\texttt{1}}
}\\
\textsuperscript{\texttt{1}}LoopLeap Lab, University of Central Florida\\
\textsuperscript{\texttt{2}}Florida State University,
\textsuperscript{\texttt{3}}Florida International University\\
\texttt{\{jiaqi.xue,yifei.zhao,qian.lou,mengxin.zheng\}@ucf.edu}\\[4pt]
{\hypersetup{pdfborder={0 0 0}}%
\small\begin{tabular}{@{}l@{\enspace}l@{}}\faGlobe~Website & \href{https://loopleap-lab.github.io/OpenMarkX/}{\color{titlelink}\texttt{https://loopleap-lab.github.io/OpenMarkX/}}\\
\faGithub~Code & \href{https://github.com/loopleap-lab/OpenMarkX}{\color{titlelink}\texttt{https://github.com/loopleap-lab/OpenMarkX}}\end{tabular}}
}

\hypersetup{pdftitle={OpenMarkX: An Extensible Platform for Watermarking Open-Source Model and Agent Outputs},pdfauthor={Jiaqi Xue, Yifei Zhao, Mansour Al Ghanim, Shangqian Gao, Ruimin Sun, Qian Lou, Mengxin Zheng}}
\begin{document}

\maketitle
\vspace{-0.2in}
\begin{figure}[H]
\centering
\vspace{-0.03in}
\includegraphics[width=\linewidth]{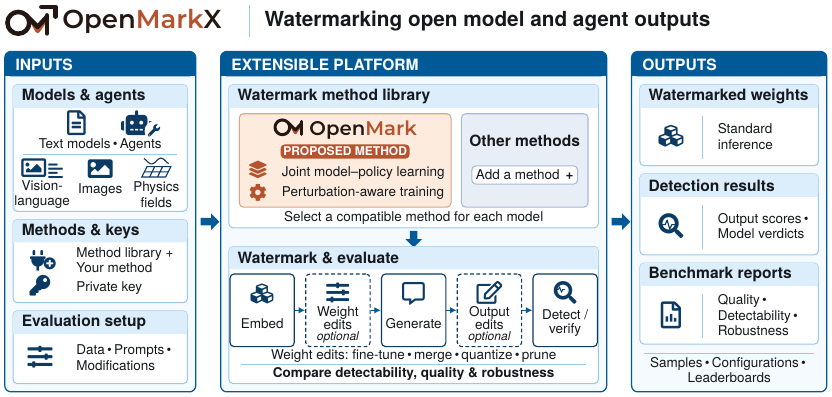}
\vspace{-0.05in}
\caption{\textbf{Overview of \oursx{}.} Configurable models and watermark methods feed embedding, generation, and evaluation. \ours{} supplies joint learning and perturbation-aware training.}
\label{fig:overview}
\end{figure}
\vspace{-0.15in}
\begin{abstract}
We introduce \oursx{}, an extensible platform for watermarking open-source model and agent outputs under standard inference. It takes models, compatible methods, private keys, and evaluation settings, and produces watermarked weights, detection results, and quality and robustness reports. Its 16 text-watermark configurations are complemented by agent, vision-language, image, and physics-field workflows. Our proposed method, \ours{}, addresses \emph{learnability} and \emph{robustness} through joint model--policy learning and perturbation-aware training. On several open-source LLMs, \ours{} achieves 0.92 TPR at 1\% FPR while reducing perplexity by 20.5\% on average relative to the strongest prior method, and remains more detectable after such modifications, e.g., retaining 0.57 TPR at 5\% FPR after 0.5-ratio model merging, compared with at most 0.34 for prior methods.
\end{abstract}

\input{sections/1_intro}
\input{sections/2_background}
\input{sections/5_platform}

\input{sections/3_method}
\input{sections/4_experiment}
\input{sections/6_conclusion}

{\section*{Acknowledgements}
This work was supported in part by the National Science Foundation under Award No. 2530786. Any opinions, findings, and conclusions or recommendations
expressed in this material are those of the author(s) and do not necessarily reflect the views of the National Science Foundation.}

{
\small
\bibliographystyle{unsrtnat}
\bibliography{citation}
}

\newpage
\appendix
\input{sections/appendix}


\end{document}

%% file: sections/1_intro.tex
\section{Introduction}


As language models enter applications~\cite{pmlr-v306-xue26h,zhao2026realm,kabir2026hw,xue2026chain,xue2026aecp,kumar2026fhecoder,kabir2026inframind}, their outputs can circulate long after generation. Concerns about misuse~\citep{bommasani2021opportunities, union2021proposal,xue2026badragidentifyingvulnerabilitiesretrieval,NEURIPS2023_cf04d01a,xue2024badfair,zheng-etal-2024-trojfsp} motivate a provenance question: can the owner of a released model recognize its outputs? A watermark embeds a statistical signal that the owner can test for later~\citep{hastuti2025factuality, ghanim2025uncovering,lou2024cr,hastuti2026evaluating}, providing evidence of model origin for attribution and IP protection~\cite{Zhao2026SIFSI,zhang2025attestllm}.

%


Figure~\ref{fig:overview} introduces \oursx{}, a platform for embedding and evaluating watermarks in open-source model and agent outputs. Here, open-source means that downstream users can access the weights and control inference. Our experiments focus on text, including an agent's final answer. We also demonstrate compatible methods on vision-language, image, and physics models.

\begin{figure*}[h!]
\centering
\includegraphics[width=0.95\textwidth]{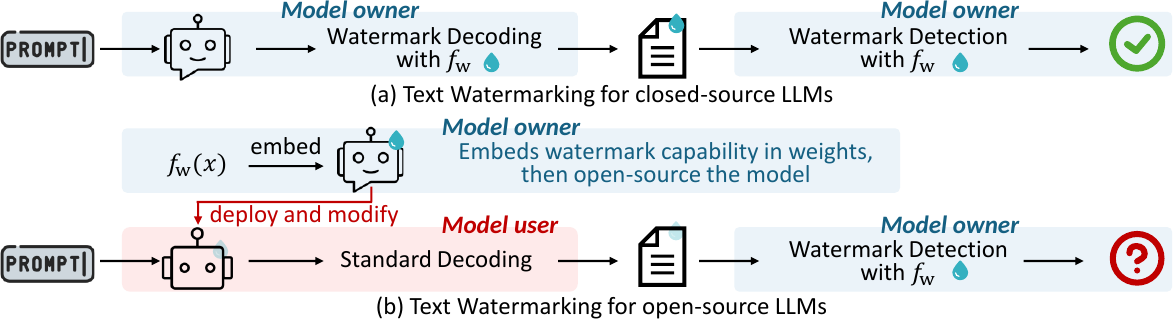}
\caption{Text watermarking for (a) closed-source and (b) open-source LLMs. Closed-source watermarking relies on watermark decoding, while open-source watermarking requires embedding the watermark into the model weights so that standard decoding still produces watermarked text.}
\label{fig:threat-model}
\end{figure*}

Figure~\ref{fig:threat-model} illustrates why access to the weights matters. In panel~(a), a user sends a prompt to a closed-source service. The provider controls decoding and uses a secret rule to favor certain tokens~\citep{kuditipudi2024robust, kirchenbauer2023watermark, hu2023unbiased}. A detector can later test the returned text for that preference. In panel~(b), the user downloads the weights and chooses the decoder. The user can omit an external watermarking step, so the owner trains the signal into the weights before release. Ordinary decoding then produces text that the owner can check.


An owner can add watermark logits to the output-layer bias~\cite{christ2024provably}, but many open models omit that bias, making the added component easy to locate and remove. Learning-based methods instead train the model to reproduce a watermarked teacher's outputs~\citep{gu2024on, sander2024watermarking}.

\begin{wrapfigure}{r}{0.52\textwidth}
\vspace{-0.15in}
\centering
\includegraphics[width=\linewidth]{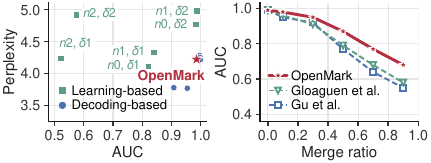}
\caption{\textit{(Left)} learning-based and decoding-based KGW watermarks under varying watermark hyperparameters \(n\) and \(\delta\). \textit{(Right)} existing learning-based watermarks' AUC after merging with the unwatermarked LLM.}
\label{fig:motivation}
\vspace{-0.1in}
\end{wrapfigure}
Figure~\ref{fig:motivation} shows two difficulties. First, a strong teacher signal is easier to detect but can reduce generation quality: increasing KGW strength from \(\delta=1\) to \(2\) raises AUC from \(0.84\) to \(0.99\) and PPL from \(4.1\) to \(5.0\). The decoding-based comparison achieves \(\text{AUC}>0.9\) with \(\text{PPL}<4.5\). Longer prefixes make the rule harder to approximate~\citep{kirchenbauer2023watermark, zhao2025can}, yet prior learning-based schemes typically use \(n\leq1\)~\citep{gu2024on, sander2024watermarking, zhao2025can}: even at \(\delta=2\), the learned 2-gram KGW watermark reaches only \(0.54\) AUC. Second, editing the released weights weakens the signal: at a merge ratio of \(0.5\), the prior method reaches \(0.79\) AUC and \ours{} retains \(0.87\).


Our method, \ours{}, jointly trains the generator and the watermark policy used for detection. This addresses a mismatch in distillation methods~\citep{gu2024on, sander2024watermarking}: a student may only partially learn a fixed rule such as KGW~\citep{kirchenbauer2023watermark}, while the detector still tests for the original rule. Joint training lets the policy adapt to patterns the generator can learn.

To improve robustness, \ours{} also trains against a weight perturbation estimated from anti-watermarked text. Earlier work selects parameters that remain stable during fine-tuning on a particular dataset~\citep{gloaguen2025towards}. Our objective directly penalizes loss of the watermark along an estimated erasure direction (Section~\ref{sec:methods}).


We compare \ours{} with prior methods~\citep{gu2024on, gloaguen2025towards} on LLaMA3-8B, Phi2-2.7B, and LLaMA3.2-3B. After 300 and 1,500 fine-tuning steps, it retains TPRs of 0.70 and 0.37 at 5\% FPR, versus baseline maxima of 0.48 and 0.23. Section~\ref{sec:openmarkx} describes the platform, and Section~\ref{sec:experiments} evaluates detection, generation quality, and robustness across edits. Table~\ref{tab:beyond-text} reports the four demonstrations beyond standalone text generation.






%% file: sections/2_background.tex
\vspace{-0.08in}
\section{Related Work and Threat Model}

\vspace{-0.06in}
\subsection{Text watermarks for Closed-source LLMs}
To ensure traceability of content generated by LLMs, watermarking techniques have been proposed to embed identifiable statistical signals into model outputs. Among these, \textit{decoding-based watermarking}~\citep{kirchenbauer2023watermark, christ2024provably, zhao2024provable, kirchenbauer2023reliability} is a widely adopted approach. The watermark decoding function $f_w(\pi_{\theta}(x), \xi)$ leverages a secret key $\xi$ to transform the original next-token distribution $\pi_{\theta}(\cdot\,|\,x)$ into a modified distribution that embeds a detectable watermark signal in the generated text. This enables post-hoc detection via a test function $f_d(x, \xi)$, which computes a p-value indicating the presence of a watermark. However, decoding-based watermarking relies on customized decoding algorithms and is not applicable in open-model settings where users control the decoding process.

\subsection{Text watermarks for Open-source LLMs}

Current watermarking schemes for open-source language models fall into two types: \textit{Input Prompt-dependent} and \textit{Input Prompt-independent}. The former requires the input prompt for detection, while the latter detects watermarks from output text alone. Notably, \ours{} is input prompt-independent.

\subsubsection{Input Prompt-dependent}
Several recent works propose open-source watermarking techniques that require access to the input prompt for detection. \cite{xu2024learning} jointly train a detector and an LLM to embed detectable signals into generated text, while \cite{block2025gaussmark} perturbs model weights with Gaussian noise and detects watermarks using gradient-based statistical tests. Both approaches rely on the input prompt during detection, which limits their practicality in real-world settings.

\subsubsection{Input Prompt-independent}
We further divide input prompt-independent watermarking methods into two classes: learning-based watermarks and structural-editing watermarks.

\noindent\textit{Learning-based Watermark.}
\cite{gu2024on} shows that decoding-based watermarks can be embedded into model weights by distilling the watermark from a teacher LLM $\pi_o$.
Then, the same watermark scheme can be used to detect the watermark in the student LLM $\pi_{\theta}$.
The teacher generates watermarked data $\mathcal{D}_{\text{wm}}$, which the student fine-tunes using the cross-entropy loss:
\begin{equation}
\small
 \mathcal{L}_{\text{sampling}} (\pi_{\theta}) = \mathbb{E}_{x \sim \mathcal{D}_{\text{wm}}} \big[ \sum_{t=1}^{|x|} -\log \pi_{\theta} (x_t | x_{<t}) \big]
\end{equation}
Alternatively, by exploiting the logits, the student can be fine-tuned to mimic the teacher LLM's next-token distribution using the KL-divergence loss:
\begin{equation}
\small
\label{eq:distillation_learnability}
 \mathcal{L}_{\text{logit}} (\pi_{\theta}) = \mathbb{E}_{x \sim \mathcal{D}} \big[ \sum_{t=1}^{|x|} \text{KL}(f_w(\pi_o(\cdot\,|\,x_{<t}), \xi_w) \parallel \pi_{\theta}(\cdot\,|\,x_{<t})) \big]
\end{equation}

\noindent\textit{Structural-editing Watermarks.}
\cite{christ2024provably} embed watermarks by adding small, token-specific Gaussian perturbations to the model's output-layer bias. 
Detection aggregates the bias perturbations of each output token. As most open-source LLMs disable output-layer biases by default\footnote{For example, Hugging Face Transformers defines the output layers of \href{https://github.com/huggingface/transformers/blob/main/src/transformers/models/llama/modeling_llama.py}{LLaMA}, \href{https://github.com/huggingface/transformers/blob/main/src/transformers/models/qwen3/modeling_qwen3.py}{Qwen}, and \href{https://github.com/huggingface/transformers/blob/main/src/transformers/models/mistral3/modeling_mistral3.py}{Mistral} with \texttt{bias=False}.}, implementing this method requires explicitly enabling the bias term, introducing an architectural modification. Such changes remain identifiable and removable by analyzing and modifying the model architecture. We therefore focus on learning-based methods as the closest comparison to \ours{}, since they share the goal of making standard decoding emit watermarked text without architectural modifications.

\subsection{Agent Output and Behavioral Watermarking}
Agent watermarking can target generated content or the behavior that produces it. AgentMark embeds identifiers in planning decisions using distribution-preserving sampling~\citep{huang2026agentmark}; ActHook embeds key-triggered hook actions in agent trajectory datasets~\citep{meng2026acthook}; and SeqWM watermarks history-conditioned behavioral transitions~\citep{an2026seqwm}. These approaches address action or trajectory provenance. The current \oursx{} agent workflow instead applies a compatible content watermark to an open language-model backbone and detects its signal in the agent's final generated answer. This scope complements behavioral watermarking and does not claim to authenticate every tool call or the entire execution trajectory.

\subsection{Threat Model}
The following threat model governs the \ours{} method and its text-generation evaluation. \oursx{} additionally records each selected method's detector-access requirements and evaluation configuration, rather than assuming that every method has the same access contract. We consider a threat model where the LLM owner publicly releases the watermarked model weights while keeping both the unwatermarked teacher $\pi_o$ and detector $M$ private~\cite{zhang2025cipherprune,santriaji2025dataseal,xue2026robpi,xue2026dictpfl,xue2024cryptotrain,yudha2024boostcom,xue2026sokfhe,xue2025twinshield,lou2023vfhe,xue2022audit}, as is standard in closed-source watermarks where the watermark seed is also private~\citep{kirchenbauer2023watermark}. The owner's goal is model-level provenance: verifying whether a given text carries the watermark of the released model. Detection is performed solely on the generated text, without access to the user's input prompt or decoding process. {With full white-box access to the released model, the user can try to evade detection at two complementary levels: (i) the \emph{text level} (paraphrasing or rewriting generated outputs, as in closed-source watermark attacks~\citep{krishna2023paraphrasing, ghanim2024jailbreaking}), and (ii) the \emph{model-weights level} (fine-tuning, merging, quantization, or pruning the released model)~\citep{lou2023trojtext, ghanim2023trojbits, cai2024wbp, zheng2024sslcleanse, zheng2024trojfair, xue2022estas}.}

%% file: sections/5_platform.tex
\vspace{-0.08in}
\section{The \oursx{} Platform}
\label{sec:openmarkx}

A run in Figure~\ref{fig:overview} starts with a model or agent backbone, a watermark method, a private key, and evaluation data. Each configuration specifies the output unit the detector can score, from a token in a final answer to a value in a predicted physical field.

\paragraph{Extending a run.}
A model adapter loads the backbone and generates outputs. A watermark module defines how to embed the signal and score those outputs, and an edit module modifies weights or generated samples. A new backbone needs a wrapper and a compatible watermark module integrated with the stages it will run. For example, the ReAct configuration tests final-answer text, whereas the physics detector compares predicted fields with a reference prediction.

\paragraph{Method and edit compatibility.}
The text benchmark has 16 configurations, including \ours{} variants, 12 decoding-time watermarks from MarkLLM~\citep{pan2024markllm} distilled into weights~\citep{gu2024on}, and GaussMark~\citep{block2025gaussmark}. Table~\ref{tab:beyond-text} includes WMAR~\citep{jovanovic2025watermarking} for image tokens and Ripple for physics fields. Each configuration selects applicable stages. A method that fails to learn a detectable signal does not complete the robustness evaluation.

\paragraph{Evaluation and outputs.}
The owner releases the watermarked weights for standard inference and retains the key and detector artifacts. A baseline run measures detectability and output quality. To test robustness, the evaluator edits weights before generation (fine-tuning, merging, quantization, or pruning) or edits outputs before detection. Run records include the configuration, samples, detection scores, model-verification results, and comparative reports.

%% file: sections/3_method.tex
\section{\ours{}: The Proposed Watermarking Method}\label{sec:methods}

This section describes \ours{}, the proposed text-watermarking method within \oursx{}. Its learned policy and detector address generation--detection consistency, while perturbation-aware training improves resistance to weight modifications. The platform's other methods retain their own embedding and detection mechanisms.

Learning-based watermarking trains an LLM to natively generate watermarked text via a teacher-student framework: applying a watermarked decoding function $f_w$ to an original LLM $\pi_o$ produces a teacher distribution $f_w(\pi_o(x), \xi)$, and a student LLM $\pi_\theta$ is trained to approximate the teacher's next-token distribution for any input prefix $x$. After training, the student generates watermarked text under standard decoding. Our goal is to build a precise and robust learning-based watermark that (i) maintains high detection accuracy without degrading generation quality, and (ii) remains robust against user modifications such as fine-tuning, merging, quantization, and pruning. We next discuss the core challenges and present how \ours{} addresses them.

CAWP and FPL address complementary sides of the learnability--robustness trade-off: CAWP co-adapts the student and watermark policy through a shared similarity objective $\mathcal{L}_{\text{sim}}$ (Eq.~\ref{eq:student_loss}) to establish a learnable signal aligned with detection, and FPL reduces the student's sensitivity to updates that erase this signal. Because both are optimized through $\mathcal{L}_{\text{sim}}$, FPL protects the signal that CAWP teaches the student.

\subsection{Co-Adaptive Watermark Policy (CAWP)}
Watermark detection for open-source LLMs involves two steps: (1) the LLM user generates text via the watermarked LLM, and (2) the LLM owner uses the predefined watermark pattern to detect the text. In this case, an important consideration is the \textit{consistency} between the watermark pattern learned by the LLM and the predefined one used during detection.
To assess this \textit{consistency}, we fine-tune a LLaMA3-8B on text generated by a decoding-based watermarked LLaMA3-8B (i.e., watermark teacher), thereby obtaining a learning-based watermarked LLaMA3-8B (i.e., watermark student). We then measure and compare the green token ratios across them in Figure~\ref{fig:motivation_all}~(left). The results reveal a significant green ratio drift in the student relative to the teacher, which reduces the AUC from \(0.99\) to \(0.84\), indicating that the student LLM fails to fully internalize the watermarking pattern of the watermark teacher.
To mitigate the inconsistency, we identify two primary optimization directions. The first is to design a learning-friendly watermark pattern, thereby enhancing its inherent learnability for the student LLM. The second is to adapt the detection mechanism itself, using a pattern that aligns more closely with the watermark distribution actually learned by the student LLM, rather than strictly adhering to the original predefined pattern.

\begin{figure*}[t!]
\centering
\includegraphics[width=0.97\textwidth]{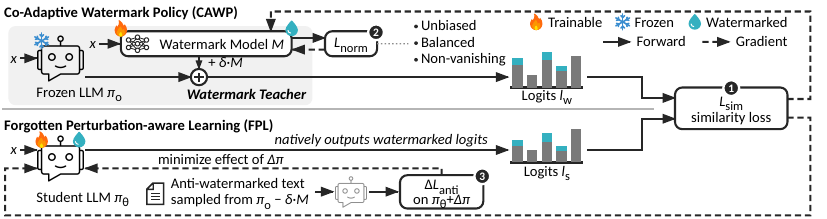}
\caption{Overview of our proposed method. CAWP (\textit{upper}) jointly trains a watermark model with the student LLM to generate learning-friendly watermark logits. FPL (\textit{bottom}) improves robustness by searching and minimizing the effect of forgotten perturbations that may erase the watermark.}
\label{fig:method}
\vspace{-0.3in}
\end{figure*}

Unlike prior work that uses rigid, predefined watermark functions, we introduce Co-Adaptive Watermark Policy (CAWP) as illustrated in Figure~\ref{fig:method}~(upper). It co-optimizes a trainable watermark policy model with the student LLM, allowing the watermark patterns to adapt to the LLM's learning dynamics and become more learnable. Detection uses the co-optimized patterns instead of the predefined ones, so it matches the watermark the LLM has internalized and avoids generation-detection inconsistency.

The watermark policy consists of a pre-trained Embedding Model \(E\) (e.g., BERT\textcolor{black}{~\citep{devlin2019bert}}) and a trainable Watermark Mapping Model \(M\) (e.g., an MLP). Given a sequence of preceding tokens \(\{x_{i-n:i-1}\}\) at position \(i\), \(E\) generates their embeddings \(\{e_{i-n:i-1}\}\). These embeddings are then transformed by \( M \) into raw watermark logits over the vocabulary, i.e., \( M(E(x_{i-n:i-1}))\). The final watermark logits are obtained by scaling with a strength coefficient \(\delta\), yielding \(\delta\cdot M(E(x_{i-n:i-1}))\), which are added to the next-token logits before sampling. Given a dataset of texts \(\mathcal{D}\), the training objective is to minimize the mean KL divergence between teacher and student next token distributions on all prefixes in \(\mathcal{D}\), shown as \lossnum{1} in Figure~\ref{fig:method}:
\begin{equation}\small
\begin{split}
\mathcal{L}_{\text{sim}}
&= \tfrac{1}{|\mathcal{D}|} \sum\nolimits_{x, i}
\mathrm{KL} \big[ \textcolor{black}{P_{\text{tea}}^{(i)}} \,\|\, \textcolor{black}{P_{\text{stu}}^{(i)}} \big], \quad
\textcolor{black}{P_{\text{stu}}^{(i)}}
= \pi_\theta(\cdot | x_{\leq i}), \\
\textcolor{black}{P_{\text{tea}}^{(i)}}
&= P\big( \log \pi_o(\cdot | x_{\leq i})
+ \delta \cdot M(E(x_{i - n:i - 1})) \big).
\end{split}
\label{eq:student_loss}
\end{equation}
Here, \(n\) is the prefix context length (in practice, a window of preceding words, e.g., 10 words, which may span a variable number of tokens) and \textcolor{black}{$P(\cdot)$ converts the logits to probabilities}. Both the teacher LLM $\pi_o$ and the embedding model $E$ are frozen; only the student model $\pi_\theta$ and mapping model $M$ are optimized. The student LLM \(\pi_\theta\) is optimized by minimizing \(\mathcal{L}_{\text{sim}}\); however, naively minimizing \(\mathcal{L}_{\text{sim}}\) for \(M\) admits degenerate policies. {Without further constraints, the joint minimum is dominated by trivial or distribution-distorting solutions, most notably \(M\!\equiv\!0\), which erases the watermark.} We therefore regularize \(M\) with three corresponding requirements, combined into $\mathcal{L}_{\text{norm}}$ and shown as \lossnum{2} in Figure~\ref{fig:method}:
(i) \textit{Unbiased Token Preference}: the watermark logits should not exhibit persistent positive or negative bias toward specific tokens. (ii) \textit{Balanced Watermark Logits}: the logits should have zero mean across the vocabulary, ensuring symmetric perturbation that makes approximately half the tokens more likely and the other half less likely. (iii) \textit{Non-Vanishing Watermark Logits}: each watermark logit is encouraged to maintain a minimum absolute magnitude, regularized toward a target value \(\epsilon\), preventing degenerate solutions where the watermark logits vanish.
\begin{equation}\small
\resizebox{\textwidth}{!}{$\displaystyle
\mathcal{L}_{\text{norm}} =
\overbrace{ \sum\nolimits_j \big| \tfrac{1}{N} \sum\nolimits_i M(e_i)^{(j)} \big| }^{\text{\scriptsize Unbiased Token Preference}}
+ \overbrace{ \sum\nolimits_i \big| \tfrac{1}{|\mathcal{V}|} \sum\nolimits_j M(e_i)^{(j)} \big| }^{\text{\scriptsize Balanced Watermark Logits}}
+ \overbrace{ \lambda_1 \sum\nolimits_{i,j} \max\big( 0, \epsilon - | M(e_i)^{(j)} | \big) }^{\text{\scriptsize Non-Vanishing Watermark Logits}}
$}
\label{eq:norm_loss_braces}
\end{equation}
\(M(e_i)^{(j)}\) denotes the watermark logit for the \(j\)-th token in the vocabulary (of size \(|\mathcal{V}|\)) corresponding to the \(i\)-th input embedding \(e_i\), with \(N\) total input samples. The index \(i\) sums over all samples, and \(j\) sums over all vocabulary tokens. To adjust the green ratio, we can replace the original \(M(e_i)^{(j)}\) inside the \textit{Unbiased Token Preference} and \textit{Balanced Watermark Logits} summations with the asymmetric weighted term. So that, for instance, to achieve a green token ratio of 0.25, we can assign a weight 3 to positive logits and 1 to negative logits; the zero-sum constraint (\(3\times N_{pos}-1\times N_{neg} \approx0\)) then compels the model to generate three times as many negative tokens as positive ones to maintain equilibrium. This weighted form lets \ours{} specify the target green-token ratio without reverting to a fixed hash table, so the learned policy can still adapt its watermark pattern to the student LLM. More technical details can be found in Appendix~\ref{app:green_ratio}.

The final training loss for the watermark mapping model \(M\) is the weighted sum of the similarity loss and the normalization loss, given by:
\begin{equation}
\label{eq:mapping_model}
\small
    \mathcal{L}_M = \mathcal{L}_{\text{sim}} + \lambda_2 \mathcal{L}_{\text{norm}}
\end{equation}
Upon training, the watermarked student LLM \(\pi_\theta\) is released, and the co-optimized watermark mapping model \(M\) is retained for detection. For a given text, the LLM owner computes the watermark logit at each position \(i\) by first embedding the \(n\)-gram prefix with \(E\), then transforming it via the mapping model \(M\) to obtain the watermark logit for the actual next token \(x_i\). The detection score is the average watermark logit across the sequence, given by \(z = \nicefrac{1}{N} \sum_{i=1}^{N} M\big(E(x_{i - n:i - 1})\big)^{(x_i)}\). {Because watermarked text is sampled from a teacher whose logits are shifted by \(\delta\cdot M(E(\cdot))\), tokens with higher \(M\) values are drawn more frequently; the average \(M\) value at the realized tokens, i.e., \(z\), is therefore systematically larger for watermarked text than for non-watermarked text.} We declare a text watermarked when $z>\tau_{\mathrm{C4}}$, where $\tau_{\mathrm{C4}}$ is calibrated once as the empirical 99th-percentile score on an independent set of 200-token human-C4 texts, disjoint from evaluation. The detector and threshold are then frozen; Appendix~\ref{app:calibration} reports realized FPR and its dependence on text length.

While some prior work also employs neural networks to generate watermark logits, they target different goals.
For example, \cite{liu2024a}, \cite{ren2023robust} and \cite{liu2023unforgeable} design semantic-invariant or public-verifiable watermark models to improve robustness or enable third-party verification. Existing approaches are designed for closed-source LLMs and do not consider the learnability of the watermark by the model itself—their watermark models are trained independently from the LLM. By contrast, CAWP focuses on generating \textit{learning-friendly} watermark logits by jointly optimizing the watermark model with the student LLM being watermarked. Additionally, such joint training ensures a bidirectional alignment: the watermark pattern used for detection adapts toward the student LLM's learned representation, while the student LLM is simultaneously guided to internalize patterns consistent with detection. This mutual convergence narrows the gap between the student and teacher distributions in Figure~\ref{fig:motivation_all}~(left).

\subsection{Forgotten Perturbation-aware Learning (FPL)}

The vulnerability of watermarked LLMs to user modifications arises primarily from \textit{forgotten perturbation}. Specifically, when a user modifies a watermarked LLM $\pi_\theta$ into $\pi'_\theta$, the weight drift $\Delta \pi_\theta = \pi'_\theta - \pi_\theta$ may remove the learned watermark. We term the drift that erases the watermark as \textit{forgotten perturbation}, and our goal is to minimize its effect during the watermark learning stage.


\begin{wrapfigure}{r}{0.52\textwidth}
\vspace{-0.15in} 
\centering
\includegraphics[width=\linewidth]{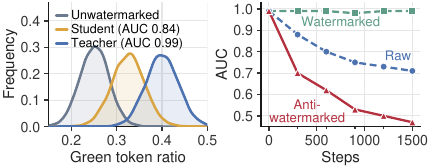}
\caption{\textit{(Left)} Green token ratios for unwatermarked, watermark student, and watermark teacher (KGW with \(\text{green ratio}=0.25\), \(n=1\), \(\delta=1\)). \textit{(Right)} AUC during fine-tuning on raw, watermarked, and anti-watermarked texts.}
\label{fig:motivation_all}
\vspace{-0.10in}
\end{wrapfigure}
To understand its root cause, we investigate fine-tuning as a representative user modification, i.e., $\pi_{t+1} = \pi_t - \eta\,g(\pi_t)$, where $g(\pi_t)$ is the gradient on user data driving weight drift. Figure~\ref{fig:motivation_all}~(right) shows that fine-tuning a watermarked LLaMA3-8B on raw texts (a mixture of green/red tokens) collapses detectability, whereas fine-tuning on watermarked texts (green-token dominant) preserves it. This indicates that gradients arising from red tokens are the primary cause of \textit{forgotten perturbation}. To further validate this, we fine-tune on \textit{anti-watermarked} texts, generated with inverted watermark logits to make texts red-token dominant. As shown in Figure~\ref{fig:motivation_all}~(right), fine-tuning on these anti-watermarked texts results in a substantial drop in detectability. These results confirm that red-token-driven \textit{forgotten perturbations} are the dominant factor in watermark forgetting, and mitigating their impact should be a key objective.

To achieve this, we propose Forgotten Perturbation-aware Learning (FPL), which {targets the worst-case erasure direction, i.e., the red-token gradient direction identified above, rather than building defenses for each user modification type}, as illustrated in Figure~\ref{fig:method}~(bottom). {We use anti-watermarked data as a proxy whose gradient \(\nabla \mathcal{L}_{\text{anti}}\) aligns with this direction, and apply a SAM-style sharpness penalty along it, so that \(\mathcal{L}_{\text{anti}}\) stays locally flat and user updates along this subspace cause minimal watermark loss.} The resulting training objective of the student LLM is:
\begin{equation}\scriptsize
\arg \min_{\bm \pi_\theta} \mathcal{L}_{\text{sim}}(\bm \pi_\theta) + \beta \big(\mathcal{L}_{\text{anti}}(\bm \pi_\theta) -  \mathcal{L}_{\text{anti}}(\bm \pi_\theta- \alpha \frac{\nabla \mathcal{L}_{\text{anti}}(\bm \pi_\theta)}{ \|\nabla \mathcal{L}_{\text{anti}}(\bm \pi_\theta) \| } ) \big)
\label{eq:booster}
\end{equation}
Here, $\mathcal{L}_{\text{sim}}$ is the watermark learning loss over watermarked texts (Eq.~\ref{eq:student_loss}), and $\mathcal{L}_{\text{anti}}$ is its analog evaluated on anti-watermarked texts (a KL divergence between the inverted teacher and the student). We denote the bracketed term by $\Delta\mathcal{L}_{\text{anti}}$, shown as \lossnum{3} in Figure~\ref{fig:method}; it is the drop of $\mathcal{L}_{\text{anti}}$ under the forgotten perturbation $\Delta\pi = -\alpha\nabla\mathcal{L}_{\text{anti}}/\|\nabla\mathcal{L}_{\text{anti}}\|$. Minimizing it keeps $\mathcal{L}_{\text{anti}}$ flat along this direction, so that the watermark is not forgotten. The hyperparameter $\beta$ controls the trade-off between watermark strength and robustness against red-token forgetting, while $\alpha$ is the perturbation step size. Optimizing Equation~\ref{eq:booster} involves second-order derivatives; a first-order approximation reduces each step to three forward/backward passes (Appendix~\ref{app:TA_4}).
\subsection{Training Procedure}\label{sec:training-procedure}
{Two modules are frozen throughout: the teacher LLM \(\pi_o\) and the embedding model \(E\); two are trainable: the student LLM \(\pi_\theta\) and the watermark mapping model \(M\). At each iteration, a minibatch updates both: \(\pi_\theta\) minimizes the FPL objective (Eq.~\ref{eq:booster}, which combines \(\mathcal{L}_{\text{sim}}\) with the forgotten-perturbation regularizer), while \(M\) minimizes \(\mathcal{L}_M = \mathcal{L}_{\text{sim}} + \lambda_2\mathcal{L}_{\text{norm}}\) (Eq.~\ref{eq:mapping_model}).} In Figure~\ref{fig:method}, $M$ is thus updated by \lossnum{1}+\lossnum{2} and $\pi_\theta$ by \lossnum{1}+\lossnum{3}.

Appendix~\ref{sec:appendix-theory} gives proxy-loss analyses of CAWP (convergence, detectability) and FPL (robustness, overhead); they motivate the design, and detection performance is evaluated empirically in Section~\ref{sec:experiments}.

%% file: sections/4_experiment.tex
\section{Evaluation}\label{sec:experiments}
We evaluate \ours{} on text generation, then assess the range of model and output types supported by \oursx{}.
\subsection{Experimental settings}

\paragraph{Models.} We perform experiments on three open-source LLMs: LLaMA3-8B, Phi2-2.7B, and LLaMA3.2-3B, which cover a range of model sizes. The watermark mapping model $M$ is a lightweight MLP: a linear projection $\mathbb{R}^{1024}\to\mathbb{R}^{500}$, two ReLU-based residual blocks, and a final projection to the vocabulary dimension with $\tanh$ to constrain logits to $[-1,1]$.

\noindent\textbf{Implementation Details.}
The embedding model and all hyperparameters are listed in Appendix~\ref{app:training-config}. All main-text \ours{} configurations use a prefix context length of \(n=10\) words on every evaluated model, and generation uses sampling with temperature 1.0 by default (other decoding settings are evaluated in Appendix~\ref{app:decoding-sweep}). Appendix~\ref{app:statistics} reports 95\% bootstrap confidence intervals and three-seed results. \ours{} adds negligible training cost: the watermark mapping model has 1.16M parameters, and FPL's extra passes are offset by faster convergence (Appendix~\ref{app:efficiency}).


\noindent\textbf{Baseline Watermarking Methods.}
Our experiments use three baseline watermarking methods, all via KL-based distillation from a decoding-based teacher model. The first is \citeauthor{gu2024on}-KGW: following~\citep{gu2024on}, we distill KGW with fixed $\gamma = 0.25$ and evaluate three $(k, \delta)$ configurations: $(1,2)$, $(0,1)$, and $(1,1)$. The second is \citeauthor{gu2024on}-KTH, which distills the exponential decoding watermark of~\citep{kuditipudi2024robust}. The third is~\citeauthor{gloaguen2025towards}-KGW, which embeds KGW watermarks into ``stable" parameters identified via contrastive task vector analysis before and after fine-tuning on raw data. Training configurations and device usage are in Appendix~\ref{appendix:modification}.

\noindent\textbf{Metrics.}
We evaluate open-source watermarking methods along three dimensions: detectability, text quality, and robustness. For detectability, following~\citep{gu2024on}, each watermarked model generates 5{,}000 samples from 50-token C4 prefixes \citep{dodge2021documenting} with 200-token continuations. Text quality is measured by median perplexity (PPL) using a {LLaMA-2-13B model following~\citep{gu2024on}}, {together with downstream reasoning accuracy on HellaSwag, ARC-Challenge, AGIEval-Math, and GSM8K to capture utility beyond distributional fluency~\citep{hastuti2026evaluating}.} To assess robustness, we apply four model modifications simulating real-world user behavior; settings are in Appendix~\ref{appendix:modification}.


\noindent\textbf{Detector calibration.} We set each detector's threshold to the 99th percentile of scores on a separate calibration set of 200-token human-written C4 texts. We freeze this threshold for evaluation. The observed false-positive rates are 1.06\% on held-out human C4 text, 1.28\% on unwatermarked outputs from the same base model, and 0.82\% on Qwen3-4B outputs (Appendix~\ref{app:calibration}).

\subsection{Results}\label{sec:results}
\begin{figure*}[h!]
\vspace{-0.1in}
    \centering
    \includegraphics[width=0.95\linewidth]{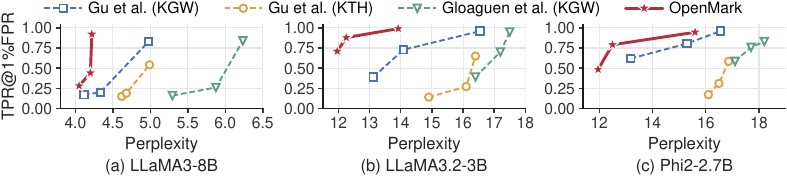}
    \caption{
Effectiveness comparison of open-source LLM text watermarks using detection TPR at 1\% FPR (TPR@1) and median PPL on three LLMs.
Better detectability and generation quality correspond to higher TPR@1 and lower PPL, i.e., lines closer to the upper left.
Original models' PPL: LLaMA3-8B = 3.6, LLaMA3.2-3B = 10.9, Phi2-2.7B = 11.3.}
\vspace{-0.1in}
    \label{fig:TPR-FPR-PPL}
\end{figure*}

\noindent\textbf{Detectability and generation quality.} Figure~\ref{fig:TPR-FPR-PPL} compares \ours{} with the methods of~\citep{gu2024on} and~\citep{gloaguen2025towards} as watermark strength varies (settings in Appendix~\ref{Appendix:baseline}). \ours{} reaches 0.92 TPR at 1\% FPR and reduces PPL by 20.5\% on average across three LLMs relative to Gu et al.-KGW, with higher TPR at 1\% FPR on each model (Appendix~\ref{app:ppl-comparison}). Figure~\ref{fig:AUC-PPL}, Table~\ref{tab:fpr_tpr}, and Figure~\ref{fig:roc} give AUC, results at 0.1\% and 10\% FPR, and ROC curves.

\begin{wraptable}{r}{0.44\textwidth}
\vspace{-0.24in}
\centering
\scriptsize
\captionsetup{font=footnotesize,skip=3pt}
\caption{Downstream task accuracy of LLaMA-3-8B before and after watermarking. \ours{} preserves capability.}
\label{tab:downstream}
\setlength{\tabcolsep}{2.7pt}\renewcommand{\arraystretch}{0.9}\setlength{\aboverulesep}{0.3ex}\setlength{\belowrulesep}{0.4ex}
\begin{tabular}{lcccc}
\toprule
\textbf{Model} & \textbf{Hella.} & \textbf{ARC-C} & \textbf{AGI-M} & \textbf{GSM8K} \\
\midrule
Base          & 0.7607 & 0.5708 & 0.2963 & 0.7583 \\
Gu et al.-KGW & 0.7583 & 0.5683 & 0.2877 & 0.7582 \\
\ours{} & 0.7594 & 0.5705 & 0.2920 & 0.7582 \\
\bottomrule
\end{tabular}
\vspace{-0.15in}
\end{wraptable}

\noindent\textbf{Downstream Task Preservation.}
{Beyond PPL, we evaluate whether watermarking degrades reasoning ability.
Table~\ref{tab:downstream} reports accuracy on four benchmarks for base LLaMA-3-8B, the KGW-watermarked variant~\citep{gu2024on}, and \ours{}.
Across the four tasks, \ours{} accuracy drops by at most 0.43 percentage points relative to the unwatermarked model. The largest drop occurs on AGIEval-Math (0.2963 to 0.2920).}

\noindent\textbf{Robustness against Model Modifications.} Following \citep{gloaguen2025towards}, we evaluate \ours{} under common model modifications, including quantization, pruning, model merging, and fine-tuning~\citep{cai2024wbp}.
As shown in Table~\ref{tab:robustness}, existing methods remain reliable under \textit{quantization} and \textit{pruning}, where parameter shifts are limited and learned weights are only mildly affected.
However, under \textit{model merging} and \textit{fine-tuning}, which cause larger and less predictable weight changes, existing methods degrade noticeably.
\ours{} degrades more slowly than all baselines under both. Merging with the original model at $t=0.5$ leaves \ours{} with TPR@5\%/1\% FPR of 0.57/0.28, versus at most 0.34/0.14 for baselines. Under OpenMathInstruct fine-tuning, \ours{} retains 0.70/0.43 after 300 steps and 0.37/0.17 after 1{,}500 steps, versus at most 0.48/0.22 and 0.23/0.08 for baselines; \ours{} also leads at every step on OpenCodeInstruct. FPL's controlled ablation and mechanism diagnostics are in Appendices~\ref{app:ablation-fpl} and~\ref{app:fpl-diagnostics}.

\noindent\textbf{Robustness against Text Paraphrasing.} \ours{} is more robust to paraphrasing attacks~\citep{krishna2023paraphrasing, ghanim2024jailbreaking} than baselines, achieving an AUC of $0.90$ under DIPPER~\citep{krishna2023paraphrasing}, compared to $0.86$ for Gu et al.-KGW (See Appendix~\ref{app:paraphrasing_attack}). Under Qwen3.5-9B paraphrasing, \ours{} retains TPR@1\% FPR of 0.32 versus 0.25 for Gu et al.-KGW after one round, and 0.17 versus 0.12 after three rounds.

\input{tables/robustness_FT_Merge}

\noindent\textbf{Short texts.} At 100 tokens on Granite-4.2-3B, \ours{} reaches 0.935 TPR at 1\% FPR using the text alone and a length-specific threshold. Appendix~\ref{app:calibration}, Table~\ref{tab:short-text-main}, compares both methods across all five lengths.

\noindent\textbf{Cross-model attribution.} Can an owner distinguish its model's outputs from those of another model carrying an \ours{} watermark? We independently watermark three models and apply each detector to every generator's outputs (Table~\ref{tab:cross-attribution}). At a nominal 1\% FPR, the detectors flag 90.9\%--92.2\% of outputs from their own models and 0.4\%--1.3\% from the other two.

\begin{table}[htbp]
\centering
\small
\caption{Cross-model attribution at a nominal 1\% FPR. Each cell gives the percentage of outputs from the column model flagged by the row detector. Bold entries pair each detector with its own model.}
\label{tab:cross-attribution}
\setlength{\tabcolsep}{8pt}
\begin{tabular}{lccc}
\toprule
& \multicolumn{3}{c}{\textbf{Generator}} \\
\cmidrule(lr){2-4}
\textbf{Detector} & LLaMA-3-8B & GPT-J-6B & Phi-2-2.7B \\
\midrule
$M_{\text{LLaMA-3-8B}}$ & \textbf{92.2\%} & 1.3\% & 0.8\% \\
$M_{\text{GPT-J-6B}}$   & 0.9\% & \textbf{90.9\%} & 1.1\% \\
$M_{\text{Phi-2-2.7B}}$ & 0.4\% & 0.7\% & \textbf{91.2\%} \\
\bottomrule
\end{tabular}
\end{table}

\noindent\textbf{Generalization to OOD Prompts.}
\ours{} trained on C4 keeps the best detectability-quality trade-off on Wikipedia prompts without retraining, and the same pipeline generalizes to GPT-J-6B (Appendix~\ref{app:cross-arch}).


\noindent\textbf{Other Experiments.}
Appendix~\ref{app:decoding-sweep} varies the decoder, and Appendices~\ref{app:modifications_tasks} and~\ref{app:ablation-fpl} vary fine-tuning duration and FPL strength. Replacing the learned policy with a fixed neural policy~\cite{liu2024a} reduces TPR at 1\% FPR from 0.92 to 0.86 and raises PPL from 4.2 to 4.6 (Appendix~\ref{app:policy-analysis}). Appendices~\ref{app:embedding_mapping}, \ref{app:detectors}, \ref{app:cross-attribution}, and~\ref{app:distillation} examine the embedding model, attribution, and distillation data scale.

\begin{table}[t]
\centering\small
\caption{Watermarking demonstrations supported by \oursx{} across model and output types, using compatible methods and adapters. TPR at 1\% FPR; quality drop is the relative change of each model's own metric from the unwatermarked to the watermarked model.}
\label{tab:beyond-text}
\setlength{\tabcolsep}{4pt}
\begin{tabular}{lllcc}
\toprule
Output & Model & Policy acts on & TPR@1\% & Quality drop \\
\midrule
Text from image & MedGemma-1.5-4B      & next-token logits         & 0.86 & 0.0\% (Acc.) \\
Agent answers   & Llama-3.1-8B (ReAct) & next-token logits         & 0.82 & 8.4\% (Acc.) \\
Image           & Taming (ImageNet)    & image-codebook tokens     & 0.90 & 4.6\% (FID) \\
Physics field   & GeoPT (aircraft)     & keyed ripple on the field & 1.00 & 0.6\% (Rel.\ $L_2$) \\
\bottomrule
\end{tabular}
\vspace{-0.22in}
\end{table}

\noindent\textbf{Platform coverage.}\label{sec:beyond-text} We examine provenance beyond standalone text in the context of model reuse and multimodal systems~\citep{li2026packcache, li2026attend, radi2026keepsake, radi2026aeroworld}. Table~\ref{tab:beyond-text} uses a frozen-policy \ours{} variant for medical answers~\citep{sellergren2025medgemma, li2026cinemr} and ReAct final answers~\citep{yao2023react, llama3modelcard}, WMAR for generated images~\citep{esser2021taming, jovanovic2025watermarking}, and Ripple for physics fields~\citep{wu2026geopt}. Across these four configurations TPR ranges from 0.82 to 1.00 at 1\% FPR.

\noindent\textbf{Watermarking agents result}. For the Llama-3.1-8B ReAct agent, the watermark policy acts on next-token logits. At a 1\% false positive rate, the detector identifies 82\% of watermarked final answers (Table~\ref{tab:beyond-text}). Task accuracy drops by 8.4\% relative to the unwatermarked agent, the largest relative quality drop in the table. This evaluation tests the final answer only; it does not test the agent's intermediate actions.

%% file: tables/robustness_FT_Merge.tex
\begin{table*}[t!]
\centering
\scriptsize
\caption{TPR at 5\%, 1\% FPR and PPL of different watermark methods of LLaMA-3-8B under model modifications. The perplexity of the original LLaMA-3-8B without watermarking is 3.6. TPR@1\% is the primary robustness metric; selected bootstrap intervals and separate three-seed results appear in Appendix~\ref{app:statistics}. We lightly highlight TPR@5\% values with green ($\ge 0.8$) and yellow ($0.5$--$0.8$) for readability.
}
\label{tab:robustness}
\setlength{\tabcolsep}{2pt}
\renewcommand{\arraystretch}{0.9}

\begin{tabular}{lllcccccccccccc}
\toprule
\multirow{3}{*}{Category} & \multirow{3}{*}{Method} & \multirow{3}{*}{Config} 
& \multicolumn{3}{c}{\cite{gu2024on}-KTH}
& \multicolumn{3}{c}{\cite{gu2024on}-KGW}
& \multicolumn{3}{c}{\cite{gloaguen2025towards}-KGW} 
& \multicolumn{3}{c}{\ours{}} \\
\cmidrule(lr){4-6} \cmidrule(lr){7-9} \cmidrule(lr){10-12} \cmidrule(lr){13-15}
&&& TPR@5 & \textbf{TPR@1} & PPL & TPR@5 & \textbf{TPR@1} & PPL & TPR@5 & \textbf{TPR@1} & PPL & TPR@5 & \textbf{TPR@1} & PPL \\
\midrule

Unaltered &&& \cellcolor{tprmid}0.78 & 0.54 & 5.0 & \cellcolor{tprgood}0.96 & 0.83 & 4.9 & \cellcolor{tprgood}0.94 & 0.84 & 6.2 & \cellcolor{tprgood}0.99 & 0.92 & 4.2 \\
\midrule

\multirow{4}{*}{Quantization}
  & \multirow{2}{*}{8 bits} 
  & \cellcolor{white}GPTQ 
  & \cellcolor{tprmid}0.66 & \cellcolor{white}0.42 & \cellcolor{white}5.2 
  & \cellcolor{tprgood}0.92 & \cellcolor{white}0.79 & \cellcolor{white}5.0 
  & \cellcolor{tprgood}0.95 & \cellcolor{white}0.84 & \cellcolor{white}6.3 
  & \cellcolor{tprgood}0.97 & \cellcolor{white}0.86 & \cellcolor{white}4.3 \\
  && \cellcolor{white}INT8 
  & \cellcolor{tprmid}0.69 & \cellcolor{white}0.45 & \cellcolor{white}5.1 
  & \cellcolor{tprgood}0.91 & \cellcolor{white}0.73 & \cellcolor{white}5.0 
  & \cellcolor{tprgood}0.94 & \cellcolor{white}0.84 & \cellcolor{white}6.4 
  & \cellcolor{tprgood}0.93 & \cellcolor{white}0.77 & \cellcolor{white}4.4 \\
  \cmidrule{2-15}
  & \multirow{2}{*}{4 bits} 
  & \cellcolor{white}HQQ 
  & \cellcolor{tprmid}0.54 & \cellcolor{white}0.29 & \cellcolor{white}6.1 
  & \cellcolor{tprgood}0.95 & \cellcolor{white}0.83 & \cellcolor{white}5.5 
  & \cellcolor{tprgood}0.91 & \cellcolor{white}0.77 & \cellcolor{white}7.9 
  & \cellcolor{tprgood}0.98 & \cellcolor{white}0.91 & \cellcolor{white}4.8 \\
  && \cellcolor{white}GPTQ 
  & \cellcolor{tprmid}0.64 & \cellcolor{white}0.36 & \cellcolor{white}5.5 
  & \cellcolor{tprgood}0.91 & \cellcolor{white}0.74 & \cellcolor{white}5.1 
  & \cellcolor{tprgood}0.94 & \cellcolor{white}0.81 & \cellcolor{white}7.3 
  & \cellcolor{tprgood}0.94 & \cellcolor{white}0.79 & \cellcolor{white}5.1 \\
\midrule

\multirow{4}{*}{Pruning}
  & \multirow{2}{*}{Wanda} 
  & \cellcolor{white}$\rho=0.2$ 
  & \cellcolor{tprmid}0.73 & \cellcolor{white}0.47 & \cellcolor{white}8.6 
  & \cellcolor{tprgood}0.96 & \cellcolor{white}0.84 & \cellcolor{white}7.5 
  & \cellcolor{tprgood}0.97 & \cellcolor{white}0.90 & \cellcolor{white}9.9 
  & \cellcolor{tprgood}0.99 & \cellcolor{white}0.92 & \cellcolor{white}7.0 \\
  && \cellcolor{white}$\rho=0.5$ 
  & \cellcolor{tprmid}0.64 & \cellcolor{white}0.35 & \cellcolor{white}8.4 
  & \cellcolor{tprgood}0.86 & \cellcolor{white}0.66 & \cellcolor{white}6.9 
  & \cellcolor{tprgood}0.93 & \cellcolor{white}0.80 & \cellcolor{white}8.9 
  & \cellcolor{tprgood}0.96 & \cellcolor{white}0.84 & \cellcolor{white}6.2 \\
  \cmidrule{2-15}
  & \multirow{2}{*}{SparseGPT} 
  & \cellcolor{white}$\rho=0.2$ 
  & \cellcolor{tprmid}0.78 & \cellcolor{white}0.55 & \cellcolor{white}7.7 
  & \cellcolor{tprgood}0.95 & \cellcolor{white}0.82 & \cellcolor{white}8.1 
  & \cellcolor{tprgood}0.95 & \cellcolor{white}0.84 & \cellcolor{white}9.0 
  & \cellcolor{tprgood}0.98 & \cellcolor{white}0.91 & \cellcolor{white}7.1 \\
  && \cellcolor{white}$\rho=0.5$ 
  & \cellcolor{tprmid}0.76 & \cellcolor{white}0.50 & \cellcolor{white}8.1 
  & \cellcolor{tprgood}0.91 & \cellcolor{white}0.74 & \cellcolor{white}7.8 
  & \cellcolor{tprgood}0.93 & \cellcolor{white}0.78 & \cellcolor{white}9.6 
  & \cellcolor{tprgood}0.97 & \cellcolor{white}0.89 & \cellcolor{white}8.0 \\
\midrule

\multirow{5}{*}{Merging}
  & \multirow{5}{*}{SLERP} 
  & \cellcolor{white}$t=0.1$ 
  & \cellcolor{white}0.36 & \cellcolor{white}0.15 & \cellcolor{white}5.2 
  & \cellcolor{tprmid}0.76 & \cellcolor{white}0.53 & \cellcolor{white}4.3 
  & \cellcolor{tprgood}0.82 & \cellcolor{white}0.59 & \cellcolor{white}6.1 
  & \cellcolor{tprgood}0.92 & \cellcolor{white}0.78 & \cellcolor{white}4.2 \\
  && \cellcolor{white}$t=0.3$ 
  & \cellcolor{white}0.21 & \cellcolor{white}0.08 & \cellcolor{white}5.1 
  & \cellcolor{tprmid}0.65 & \cellcolor{white}0.39 & \cellcolor{white}4.0 
  & \cellcolor{tprmid}0.62 & \cellcolor{white}0.37 & \cellcolor{white}5.3 
  & \cellcolor{tprgood}0.83 & \cellcolor{white}0.60 & \cellcolor{white}4.1 \\
  && \cellcolor{white}$t=0.5$ 
  & \cellcolor{white}0.09 & \cellcolor{white}0.02 & \cellcolor{white}4.8 
  & \cellcolor{white}0.30 & \cellcolor{white}0.12 & \cellcolor{white}3.8 
  & \cellcolor{white}0.34 & \cellcolor{white}0.14 & \cellcolor{white}4.6 
  & \cellcolor{tprmid}0.57 & \cellcolor{white}0.28 & \cellcolor{white}4.0 \\
  && \cellcolor{white}$t=0.7$ 
  & \cellcolor{white}0.06 & \cellcolor{white}0.02 & \cellcolor{white}4.3 
  & \cellcolor{white}0.15 & \cellcolor{white}0.05 & \cellcolor{white}3.7 
  & \cellcolor{white}0.19 & \cellcolor{white}0.06 & \cellcolor{white}4.2 
  & \cellcolor{white}0.31 & \cellcolor{white}0.12 & \cellcolor{white}3.9 \\
  && \cellcolor{white}$t=0.9$ 
  & \cellcolor{white}0.06 & \cellcolor{white}0.01 & \cellcolor{white}4.1 
  & \cellcolor{white}0.11 & \cellcolor{white}0.03 & \cellcolor{white}3.6 
  & \cellcolor{white}0.11 & \cellcolor{white}0.03 & \cellcolor{white}4.0 
  & \cellcolor{white}0.17 & \cellcolor{white}0.06 & \cellcolor{white}3.8 \\
\midrule

\multirow{10}{*}{Finetuning}
  & \multirow{4}{*}{\makecell[l]{OpenMath\\Instruct}}
  & \cellcolor{white}\(s=300\)  
  & \cellcolor{white}0.21 & \cellcolor{white}0.08 & \cellcolor{white}5.1 
  & \cellcolor{white}0.48 & \cellcolor{white}0.20 & \cellcolor{white}4.1 
  & \cellcolor{white}0.45 & \cellcolor{white}0.22 & \cellcolor{white}5.5 
  & \cellcolor{tprmid}0.70 & \cellcolor{white}0.43 & \cellcolor{white}4.1 \\
  && \cellcolor{white}\(s=600\) 
  & \cellcolor{white}0.17 & \cellcolor{white}0.05 & \cellcolor{white}4.9 
  & \cellcolor{white}0.38 & \cellcolor{white}0.16 & \cellcolor{white}3.7 
  & \cellcolor{white}0.41 & \cellcolor{white}0.19 & \cellcolor{white}5.2 
  & \cellcolor{tprmid}0.51 & \cellcolor{white}0.26 & \cellcolor{white}3.9 \\
  && \cellcolor{white}\(s=900\) 
  & \cellcolor{white}0.09 & \cellcolor{white}0.03 & \cellcolor{white}4.8 
  & \cellcolor{white}0.30 & \cellcolor{white}0.11 & \cellcolor{white}3.6 
  & \cellcolor{white}0.31 & \cellcolor{white}0.13 & \cellcolor{white}5.2 
  & \cellcolor{white}0.46 & \cellcolor{white}0.22 & \cellcolor{white}3.7 \\
  && \cellcolor{white}\(s=1200\) 
  & \cellcolor{white}0.09 & \cellcolor{white}0.02 & \cellcolor{white}4.7 
  & \cellcolor{white}0.25 & \cellcolor{white}0.09 & \cellcolor{white}3.5 
  & \cellcolor{white}0.25 & \cellcolor{white}0.10 & \cellcolor{white}5.0 
  & \cellcolor{white}0.40 & \cellcolor{white}0.18 & \cellcolor{white}3.6 \\
  && \cellcolor{white}\(s=1500\) 
  & \cellcolor{white}0.02 & \cellcolor{white}0.02 & \cellcolor{white}4.5 
  & \cellcolor{white}0.23 & \cellcolor{white}0.08 & \cellcolor{white}3.5 
  & \cellcolor{white}0.22 & \cellcolor{white}0.08 & \cellcolor{white}5.1 
  & \cellcolor{white}0.37 & \cellcolor{white}0.17 & \cellcolor{white}3.6 \\
  \cmidrule{2-15}
  & \multirow{4}{*}{\makecell[l]{OpenCode \\ Instruct}}
  & \cellcolor{white}\(s=300\) 
  & \cellcolor{white}0.27 & \cellcolor{white}0.11 & \cellcolor{white}7.3 
  & \cellcolor{tprmid}0.63 & \cellcolor{white}0.40 & \cellcolor{white}6.9 
  & \cellcolor{tprmid}0.65 & \cellcolor{white}0.41 & \cellcolor{white}7.4 
  & \cellcolor{tprmid}0.71 & \cellcolor{white}0.43 & \cellcolor{white}7.0 \\
  && \cellcolor{white}\(s=600\) 
  & \cellcolor{white}0.22 & \cellcolor{white}0.08 & \cellcolor{white}5.7 
  & \cellcolor{white}0.45 & \cellcolor{white}0.24 & \cellcolor{white}4.9 
  & \cellcolor{white}0.48 & \cellcolor{white}0.23 & \cellcolor{white}7.0 
  & \cellcolor{tprmid}0.56 & \cellcolor{white}0.31 & \cellcolor{white}5.4 \\
  && \cellcolor{white}\(s=900\) 
  & \cellcolor{white}0.16 & \cellcolor{white}0.06 & \cellcolor{white}4.8 
  & \cellcolor{white}0.32 & \cellcolor{white}0.14 & \cellcolor{white}5.0 
  & \cellcolor{white}0.34 & \cellcolor{white}0.16 & \cellcolor{white}6.8 
  & \cellcolor{tprmid}0.50 & \cellcolor{white}0.25 & \cellcolor{white}5.1 \\
  && \cellcolor{white}\(s=1200\) 
  & \cellcolor{white}0.11 & \cellcolor{white}0.03 & \cellcolor{white}4.7 
  & \cellcolor{white}0.29 & \cellcolor{white}0.12 & \cellcolor{white}4.3 
  & \cellcolor{white}0.31 & \cellcolor{white}0.12 & \cellcolor{white}6.3 
  & \cellcolor{white}0.40 & \cellcolor{white}0.18 & \cellcolor{white}4.8 \\
  && \cellcolor{white}\(s=1500\) 
  & \cellcolor{white}0.07 & \cellcolor{white}0.02 & \cellcolor{white}4.6 
  & \cellcolor{white}0.26 & \cellcolor{white}0.10 & \cellcolor{white}4.2 
  & \cellcolor{white}0.30 & \cellcolor{white}0.12 & \cellcolor{white}5.9 
  & \cellcolor{white}0.39 & \cellcolor{white}0.17 & \cellcolor{white}4.0 \\
\bottomrule
\end{tabular}
\end{table*}

%% file: sections/6_conclusion.tex
\vspace{-0.08in}
\section{Conclusion}
\oursx{} lets a model owner select a compatible watermark and evaluate its signal after editing the released weights or outputs. Across three text models, \ours{} reduces perplexity by 20.5\% on average relative to Gu et al.-KGW and retains 0.57 TPR at 5\% FPR after a 0.5-ratio merge, versus at most 0.34 for the baselines. The platform also supports evaluation on agent answers, medical answers, images, and physical fields.

%% file: sections/appendix.tex
\section*{Appendix Contents}
\addcontentsline{toc}{section}{Appendix Contents}

\renewcommand{\arraystretch}{1.15}
\noindent\begin{tabular}{@{}p{0.9\textwidth}@{}r@{}}
\multicolumn{2}{@{}l@{}}{\textbf{A.\quad Theoretical Analysis}}\\[1pt]
\quad A.1\quad Convergence Analysis of CAWP & \pageref{app:TA_1}\\
\quad A.2\quad Formal Detectability Advantage of CAWP over Fixed Mappings & \pageref{app:TA_2}\\
\quad A.3\quad Local Proxy-Loss Robustness of FPL & \pageref{sec:fpl-theory}\\
\quad A.4\quad Computational Overhead of FPL & \pageref{app:TA_4}\\[4pt]

\multicolumn{2}{@{}l@{}}{\textbf{B.\quad Method Background}}\\[1pt]
\quad B.1\quad Configurable Watermark Token Ratio $\gamma$ & \pageref{app:green_ratio}\\
\quad B.2\quad Closed-source Watermark Schemes (KGW \& KTH) & \pageref{app:closed-source-bg}\\
\quad B.3\quad Model Modification Techniques & \pageref{app:modification-bg}\\[4pt]

\multicolumn{2}{@{}l@{}}{\textbf{C.\quad Experimental Configuration}}\\[1pt]
\quad C.1\quad Baseline Setting & \pageref{Appendix:baseline}\\
\quad C.2\quad Training Configurations & \pageref{app:training-config}\\
\quad C.3\quad Model Modification Settings & \pageref{appendix:modification}\\[4pt]

\multicolumn{2}{@{}l@{}}{\textbf{D.\quad Additional Empirical Results}}\\[1pt]
\quad D.1\quad Computational Efficiency Analysis & \pageref{app:efficiency}\\
\quad D.2\quad Cross-Distribution and Cross-Architecture Generalization & \pageref{app:cross-arch}\\
\quad D.3\quad Ablation on FPL Hyperparameters & \pageref{app:ablation-fpl}\\
\quad D.4\quad Visualization of ROC Curves & \pageref{app:ROC}\\
\quad D.5\quad Detection Performance under Different FPRs & \pageref{app:fpr_tpr}\\
\quad D.6\quad Overall Detection AUC & \pageref{app:AUC}\\
\quad D.7\quad Robustness against Paraphrasing Attack & \pageref{app:paraphrasing_attack}\\
\quad D.8\quad Robustness across Downstream Task Modifications & \pageref{app:modifications_tasks}\\
\quad D.9\quad Comparison with Classifier-based LLM Text Detectors & \pageref{app:detectors}\\
\quad D.10\;\,Cross-Model Attribution across \ours{}-Watermarked Models & \pageref{app:cross-attribution}\\
\quad D.11\;\,Different Embedding and Mapping Models & \pageref{app:embedding_mapping}\\
\quad D.12\;\,Distillation Data Scaling Study & \pageref{app:distillation}\\
\quad D.13\;\,Null Calibration and Text Length & \pageref{app:calibration}\\
\quad D.14\;\,Statistical Uncertainty and Independent Training Runs & \pageref{app:statistics}\\
\quad D.15\;\,Decoding Hyperparameter Sweep & \pageref{app:decoding-sweep}\\
\quad D.16\;\,Learned Policy Analysis and Fixed Neural Baseline & \pageref{app:policy-analysis}\\
\quad D.17\;\,FPL Alignment and Detector-Refitting Diagnostics & \pageref{app:fpl-diagnostics}\\[4pt]

\multicolumn{2}{@{}l@{}}{\textbf{E.\quad Limitations} \dotfill \pageref{app:limitations}}\\[4pt]

\multicolumn{2}{@{}l@{}}{\textbf{F.\quad Impact Statement} \dotfill \pageref{app:impact}}\\
\end{tabular}
\vspace{1em}

\section{Theoretical Analysis}\label{sec:appendix-theory}
\subsection{Convergence Analysis of CAWP}\label{app:TA_1}

We anchor our proof in established convergence guarantees for distillation-based LLM training, where a student LLM $\pi_\theta$ converges to a fixed teacher $\pi_o$ under KL divergence minimization. The core innovation of CAWP is introducing a trainable watermark policy $M_\phi$ that perturbs the teacher distribution. We prove convergence by analyzing this perturbation.

\subsubsection{Baseline Distillation Convergence (Fixed Policy $M$)}
For a fixed watermark policy $M_{\text{fix}}$, the loss reduces to standard distillation:

\begin{equation}
\mathcal{L}_{\text{fix}}(\theta) = \text{KL}\left( \pi_o + \delta M_{\text{fix}} \parallel \pi_\theta \right).
\end{equation}

Under Lipschitz continuity and bounded gradients, gradient descent on $\theta$ converges:

\begin{equation}
\lim_{t \to \infty} \|\nabla_\theta \mathcal{L}_{\text{fix}}\| = 0.
\end{equation}

This serves as the baseline we extend.

\subsubsection{CAWP's Joint Optimization}
CAWP introduces a trainable policy $M_\phi$, leading to the joint loss:

\begin{equation}
\mathcal{L}_{\text{sim}}(\theta, \phi) =
\underbrace{\text{KL}\left( \pi_o + \delta M_\phi \parallel \pi_\theta \right)}_{\text{Distillation loss}}
+ \lambda_2 \underbrace{\mathcal{L}_{\text{norm}}(\phi)}_{\text{Regularizer}}.
\end{equation}

We analyze the effect of $M_\phi$ on convergence. Specifically, $\mathcal{L}_{\text{sim}}(\theta, \phi)$ satisfies the following assumptions:

\begin{itemize}
    \item \textbf{Smoothness:} $\nabla \text{KL}(\cdot \parallel \pi_\theta)$ is $L_\theta$-Lipschitz in $\theta$, and $\nabla M_\phi$ is $L_\phi$-Lipschitz in $\phi$. LLMs and MLPs with smooth activations (e.g., GELU, Tanh) satisfy this.
    \item \textbf{Boundedness:} $\|M_\phi\| \leq B_M$ and $\|\nabla_\phi M_\phi\| \leq G_M$, since outputs are bounded by the last Tanh layer and gradients are bounded via clipping.
    \item \textbf{Convexity:} $\mathcal{L}_{\text{norm}}$ is convex in $\phi$, as $\ell_1$ penalties in Eq.~(4) enforce convexity.
\end{itemize}

Given these assumptions, alternating gradient descent on $\theta$ and $\phi$ converges to a stationary point, where $M_\phi$ learns watermark mappings that perturb $\pi_o$ without degrading the LLM's performance.

\subsubsection{Alternating Gradient Descent Dynamics}
In CAWP, optimization alternates between updating $\theta$ (distilling to the perturbed teacher) and $\phi$ (adapting the watermark policy):

\begin{align}
\theta^{t+1} &= \theta^t - \eta_\theta \nabla_\theta \mathcal{L}_{\text{sim}}(\theta^t, \phi^t), \\
\phi^{t+1} &= \phi^t - \eta_\phi \nabla_\phi \mathcal{L}_{\text{sim}}(\theta^{t+1}, \phi^t).
\end{align}

The distillation term encourages $\pi_\theta$ to match the perturbed distribution, while $\mathcal{L}_{\text{norm}}(\phi)$ (e.g., $\ell_1$ on policy outputs) prevents $M_\phi$ from over-perturbing, ensuring watermark detectability without degrading text quality.

Under the smoothness assumption, the loss is $L$-smooth overall ($L = L_\theta + \delta L_\phi B_M$). Hence,

\begin{equation}
\begin{aligned}
\mathcal{L}_{\text{sim}}(\theta^{t+1}, \phi^{t+1}) \leq 
&\; \mathcal{L}_{\text{sim}}(\theta^t, \phi^t) 
- \Bigg( \frac{\eta_\theta}{2} \|\nabla_\theta \mathcal{L}_{\text{sim}}\|^2 
   + \frac{\eta_\phi}{2} \|\nabla_\phi \mathcal{L}_{\text{sim}}\|^2 \Bigg) \\
&+ \frac{L \eta_\theta^2}{2} \|\nabla_\theta \mathcal{L}_{\text{sim}}\|^2
  + \frac{L \eta_\phi^2}{2} \|\nabla_\phi \mathcal{L}_{\text{sim}}\|^2 .
\end{aligned}
\end{equation}

Choosing $\eta_\theta, \eta_\phi \leq 1/L$ ensures monotonic decrease:

\begin{equation}
\mathcal{L}_{\text{sim}}(\theta^{t+1}, \phi^{t+1})
\leq \mathcal{L}_{\text{sim}}(\theta^t, \phi^t) - c \Big(
\|\nabla_\theta \mathcal{L}_{\text{sim}}\|^2 + \|\nabla_\phi \mathcal{L}_{\text{sim}}\|^2
\Big), \quad c > 0.
\end{equation}

\subsubsection{Convergence to Stationary Point}
Summing the descent inequality over $T$ iterations gives:

\begin{equation}
\sum_{t=0}^{T-1} \left( \|\nabla_\theta \mathcal{L}_{\text{sim}}(\theta^t, \phi^t)\|^2 +
\|\nabla_\phi \mathcal{L}_{\text{sim}}(\theta^t, \phi^t)\|^2 \right)
\leq \frac{\mathcal{L}_{\text{sim}}(\theta^0, \phi^0) - \mathcal{L}_{\text{sim}}^*}{cT},
\end{equation}

where $\mathcal{L}_{\text{sim}}^*$ is the infimum of the loss. As $T \to \infty$, the gradients vanish:

\begin{equation}
\lim_{t \to \infty} \|\nabla_\theta \mathcal{L}_{\text{sim}}\| = 0, \quad
\lim_{t \to \infty} \|\nabla_\phi \mathcal{L}_{\text{sim}}\| = 0.
\end{equation}

The boundedness of $M_\phi$ and its gradients ensures that the perturbation $\delta M_\phi$ remains controlled, preventing divergence. Moreover, the $\mu$-strong convexity of $\mathcal{L}_{\text{norm}}$ implies that, for fixed $\theta$, the subproblem in $\phi$ is strongly convex, guaranteeing convergence to a unique minimizer $\phi^*(\theta)$ that balances watermark strength and regularization.

\subsection{Formal Detectability Advantage of CAWP over Fixed Watermark Mappings}
\label{app:TA_2}
In this section, we formally show that CAWP achieves a strictly smaller generation-detection inconsistency compared to the baseline (fixed watermark mappings). Our analysis leverages (i) the irreducible non-approximability of hash-like fixed mappings and (ii) standard generalization bounds for the joint hypothesis class of student LLMs and adaptive watermark policies.

\noindent\textbf{Notation.} Let $\pi_\theta(y\mid x)$ denote the student LLM, and let $q_{\text{fix}}(y\mid x)$ denote the teacher distribution generated by a fixed watermark mapping $M_{\text{fix}}$ (such as KGW/KTH). For CAWP, let $q_\phi(y\mid x)$ denote the teacher distribution generated by the trainable watermark policy $M_\phi$.

Define the population KL risks:
\begin{align}
  \mathcal{R}_{\text{fix}}(\theta)
  &= \mathbb{E}_{x}\,
     \mathrm{KL}\big(q_{\text{fix}}(\cdot\mid x)
     \,\Vert\, \pi_\theta(\cdot\mid x)\big), \\
  \mathcal{R}_{\text{CAWP}}(\theta,\phi)
  &= \mathbb{E}_{x}\,
     \mathrm{KL}\big(q_{\phi}(\cdot\mid x)
     \,\Vert\, \pi_\theta(\cdot\mid x)\big).
\end{align}

The optimal risks for the two settings are:
\begin{align}
  \gamma := \inf_{\theta}\mathcal{R}_{\text{fix}}(\theta), \qquad
  \varepsilon := \inf_{\theta,\phi}\mathcal{R}_{\text{CAWP}}(\theta,\phi).
\end{align}

\noindent\textbf{Assumption 1 (Irreducible error of baseline fixed mappings).} The fixed watermark mapping $M_{\text{fix}}$ is hash-like, discontinuous, and not aligned with the transformer parameterization. Thus, there exists a constant $\gamma_0>0$ such that
\begin{equation}
  \forall\,\theta:\quad
  \mathcal{R}_{\text{fix}}(\theta)
  \ge \gamma_0,
\end{equation}
and hence:
\begin{equation}
  \gamma = \inf_{\theta}\mathcal{R}_{\text{fix}}(\theta)
  \;\ge\; \gamma_0.
\end{equation}
This corresponds to the empirical mismatch observed in Figure~\ref{fig:motivation_all}~(left), where the student fails to reproduce the fixed green-token distribution.

\noindent\textbf{Assumption 2 (Generalization of CAWP).}
Let $\widehat{\mathcal{R}}_{\text{CAWP}}(\theta,\phi)$ denote the empirical CAWP loss over $N$ training samples:
\begin{equation}
  \widehat{\mathcal{R}}_{\text{CAWP}}(\theta,\phi)
  = \frac{1}{N}\sum_{i=1}^N
    \mathrm{KL}\big(q_\phi(\cdot\mid x_i)\Vert\pi_\theta(\cdot\mid x_i)\big).
\end{equation}
We assume that the joint hypothesis class $\mathcal{H}=\{(\theta,\phi)\}$ has finite capacity (e.g., finite Rademacher complexity), following standard generalization analyses of deep neural networks~\citep{bartlett2002rademacher, neyshabur2015norm, bartlett2017spectrally}. We do not require the explicit value of $\mathfrak{C}$, only that it is finite, which holds for bounded-parameter networks such as the CAWP policy and the student LLM. Moreover, the KL loss is bounded for all softmax-based language models, as both $q_\phi(\cdot\mid x)$ and $\pi_\theta(\cdot\mid x)$ are strictly positive distributions over a finite vocabulary. 

Under these standard conditions, classical uniform-convergence bounds \citep{bartlett2002rademacher} imply that there exists a constant $C>0$ such that, with probability at least $1-\delta$,
\begin{equation}
  \mathcal{R}_{\text{CAWP}}(\theta,\phi)
  \le
  \widehat{\mathcal{R}}_{\text{CAWP}}(\theta,\phi)
  +
  C\sqrt{\frac{\mathfrak{C}+\log(1/\delta)}{N}}.
  \label{eq:gen-bound}
\end{equation}

\noindent\textbf{Assumption 3 (Effective optimization of CAWP).} Let $(\theta^\star,\phi^\star)$ be an approximate empirical minimizer:
\begin{equation}
  \widehat{\mathcal{R}}_{\text{CAWP}}(\theta^\star,\phi^\star)
  \le
  \inf_{\theta,\phi}\widehat{\mathcal{R}}_{\text{CAWP}}(\theta,\phi)
  + \eta,
\end{equation}
where $\eta$ is negligible.
Because $M_\phi$ adapts to the representational capacity of the
student LLM, we may assume:
\begin{equation}
  \inf_{\theta,\phi}\widehat{\mathcal{R}}_{\text{CAWP}}(\theta,\phi)
  \approx 0,
  \qquad
  \widehat{\mathcal{R}}_{\text{CAWP}}(\theta^\star,\phi^\star)
  \approx 0.
\end{equation}

\begin{theorem}[Detectability gain of CAWP]
\label{thm:cawp-bound}
Under Assumptions~1--3, with probability at least $1-\delta$,
the optimal risks satisfy
\begin{equation}
\gamma \;\ge\; \gamma_0
\qquad\text{and}\qquad
\varepsilon \;\le\;
C\sqrt{\frac{\mathfrak{C}+\log(1/\delta)}{N}}.
\end{equation}

Consequently, the relative and absolute gaps between CAWP and fixed
watermark mappings are bounded by
\begin{equation}
\frac{\varepsilon}{\gamma}
\le
\frac{C}{\gamma_0}\sqrt{\frac{\mathfrak{C}+\log(1/\delta)}{N}}
\qquad\text{and}\qquad
\gamma-\varepsilon
\ge
\gamma_0 - C\sqrt{\frac{\mathfrak{C}+\log(1/\delta)}{N}}.
\end{equation}

\end{theorem}

The proof is detailed:

\noindent\textit{Step 1: Lower bound on $\gamma$.}
Immediate from Assumption~1.

\noindent\textit{Step 2: Upper bound on $\varepsilon$.}
Applying the generalization bound Equation~\ref{eq:gen-bound} to 
$(\theta^\star,\phi^\star)$ gives
\begin{equation}
  \mathcal{R}_{\text{CAWP}}(\theta^\star,\phi^\star)
  \le
  \widehat{\mathcal{R}}_{\text{CAWP}}(\theta^\star,\phi^\star)
  +
  C\sqrt{\frac{\mathfrak{C}+\log(1/\delta)}{N}}.
\end{equation}
Using Assumption~3 yields
\begin{equation}
  \mathcal{R}_{\text{CAWP}}(\theta^\star,\phi^\star)
  \le
  C\sqrt{\frac{\mathfrak{C}+\log(1/\delta)}{N}}.
\end{equation}
Since $\varepsilon$ is the infimum over all $(\theta,\phi)$,
\begin{equation}
  \varepsilon
  \le
  \mathcal{R}_{\text{CAWP}}(\theta^\star,\phi^\star)
  \le
  C\sqrt{\frac{\mathfrak{C}+\log(1/\delta)}{N}}.
\end{equation}
\noindent\textit{Step 3: Relative and absolute gaps.}
Combining $\gamma\ge\gamma_0$ and the above yields
\begin{equation}
  \frac{\varepsilon}{\gamma}
  \le
  \frac{C}{\gamma_0}
  \sqrt{\frac{\mathfrak{C}+\log(1/\delta)}{N}}.
\end{equation}
Furthermore,
\begin{equation}
  \gamma-\varepsilon
  \ge
  \gamma_0
  -
  C\sqrt{\frac{\mathfrak{C}+\log(1/\delta)}{N}}.
\end{equation}
This completes the proof.

\noindent\textbf{Discussion.} The bound above shows that the detectability gain of CAWP over fixed watermark mappings is driven by three intuitive factors: (i) a larger irreducible error $\gamma_0$ of the fixed mapping (harder-to-learn hash-like patterns), (ii) a larger training sample size $N$, which shrinks the generalization term, and (iii) a smaller effective capacity $\mathfrak{C}$ of the CAWP model. In other words, CAWP yields the largest improvement precisely in the regime where fixed mappings are intrinsically hard to learn, while CAWP is trained with sufficient data and a well-regularized architecture.

This comparison characterizes the population teacher--student KL mismatch and motivates CAWP's design; detection performance is evaluated empirically in Section~\ref{sec:experiments}.

\subsection{Local Proxy-Loss Robustness of FPL}
\label{sec:fpl-theory}

The analysis below compares a lower bound on the baseline's worst-case forgetting with an upper bound on FPL's, yielding a local robustness guarantee for the anti-watermark proxy loss within a bounded perturbation radius.

\noindent\textbf{Robustness setting.}
Let $\theta$ denote the parameters of a watermarked student model, and let
$\theta' = \theta + \Delta$ be a user-modified model after some 
fine-tuning, merging, or other downstream adaptation.
We assume the user perturbation is bounded in parameter space,
$\|\Delta\|\le R$, where $R>0$ is a given radius (e.g., induced by model
merging or a limited number of fine-tuning steps).

Let $q_{\mathrm{anti}}(\cdot\mid x)$ denote the \emph{anti-watermark teacher}
distribution, obtained by subtracting the green-list bias from the original
watermark logits (cf.\ Eq.~(\ref{eq:booster}) in the main text). 
We define the \emph{anti-watermark loss}
\begin{equation}
  L_{\mathrm{anti}}(\theta)
  := \mathbb{E}_{x\sim D_{\mathrm{anti}}}
  \Big[
    \mathrm{KL}\big(q_{\mathrm{anti}}(\cdot\mid x)\,\Vert\,\pi_\theta(\cdot\mid x)\big)
  \Big],
\end{equation}
where $D_{\mathrm{anti}}$ is a distribution of ``anti-watermark'' inputs
(e.g., adversarial prompts or samples generated using the anti-watermark
logits). A smaller value of $L_{\mathrm{anti}}(\theta)$ means that the
student model behaves more like the anti-watermark teacher on these
hard inputs, i.e., the embedded watermark is weaker. 
Thus ``watermark forgetting'' corresponds to a \emph{decrease} in 
$L_{\mathrm{anti}}$.

For a given radius $R$, we therefore measure robustness by the worst-case
\emph{drop} of $L_{\mathrm{anti}}$ under bounded perturbations:
\begin{equation}
  \Delta^{\downarrow}_{\max}(\theta; R)
  :=
  \max_{\|\Delta\|\le R}
  \Big\{
    L_{\mathrm{anti}}(\theta) - L_{\mathrm{anti}}(\theta+\Delta)
  \Big\}.
\end{equation}
A model with smaller $\Delta^{\downarrow}_{\max}(\theta; R)$ is more robust,
because no admissible user modification can significantly push the model
towards the anti-watermark distribution.

\medskip
\noindent\textbf{Assumption 4 (Smoothness of the anti-watermark loss).}
$L_{\mathrm{anti}}(\theta)$ is $L$-smooth, i.e., its gradient is 
Lipschitz continuous with constant $L>0$:
\begin{equation}
  \big\|\nabla L_{\mathrm{anti}}(\theta_1) 
       - \nabla L_{\mathrm{anti}}(\theta_2)\big\|
  \le
  L \,\|\theta_1-\theta_2\|
  \quad
  \forall\,\theta_1,\theta_2.
\end{equation}
This is a standard assumption in optimization for deep networks.

\medskip
\noindent\textbf{Assumption 5 (FPL reduces the forgetting gradient).}
Let $\theta_{\mathrm{base}}$ be the parameters obtained \emph{without} FPL
(e.g., using CAWP or a baseline watermarking scheme), and let 
$\theta_{\mathrm{FPL}}$ be the parameters obtained \emph{with} FPL.
Define
\begin{equation}
  G_{\mathrm{base}}
  := \big\|\nabla L_{\mathrm{anti}}(\theta_{\mathrm{base}})\big\|,
  \qquad
  G_{\mathrm{FPL}}
  := \big\|\nabla L_{\mathrm{anti}}(\theta_{\mathrm{FPL}})\big\|.
\end{equation}
Since FPL explicitly penalizes moving the model towards the anti-watermark
teacher along the worst forgetting direction (see Eq.~(\ref{eq:booster})),
it encourages solutions where the anti-watermark loss is locally flat.
We therefore assume that for some $\kappa\in(0,1)$ we have
\begin{equation}
  G_{\mathrm{FPL}} \;\le\; \kappa\,G_{\mathrm{base}},
\end{equation}
i.e., FPL reduces the norm of the anti-watermark gradient by a multiplicative
factor $\kappa$ compared to the baseline. This assumption motivates our empirical robustness tests under fine-tuning and merging, where FPL-trained models retain stronger detectability after downstream modifications.

\begin{theorem}[Local anti-watermark proxy robustness]
\label{thm:fpl-robustness}
Under Assumptions~4--5, suppose that the same smoothness constant $L$ applies throughout the radius-$R$ neighborhoods of both checkpoints and that $G_{\mathrm{base}}>0$. For any checkpoint $\theta$ in this comparison, writing $G(\theta)=\|\nabla L_{\mathrm{anti}}(\theta)\|$, we have
\begin{equation}
 RG(\theta)-\frac{L}{2}R^2
 \le \Delta^{\downarrow}_{\max}(\theta;R)
 \le RG(\theta)+\frac{L}{2}R^2.
 \label{eq:fpl-two-sided}
\end{equation}
In particular,
\begin{align}
  \Delta^{\downarrow}_{\max}(\theta_{\mathrm{base}}; R)
  &\;\ge\;
  R\,G_{\mathrm{base}} - \frac{L}{2}R^2,\\[2pt]
  \Delta^{\downarrow}_{\max}(\theta_{\mathrm{FPL}}; R)
  &\;\le\;
  R\,G_{\mathrm{FPL}} + \frac{L}{2}R^2
  \;\le\;
  R\,\kappa G_{\mathrm{base}} + \frac{L}{2}R^2.
\end{align}
Subtracting the FPL upper bound from the baseline lower bound yields
\begin{equation}
  \Delta^{\downarrow}_{\max}(\theta_{\mathrm{base}}; R)
  - \Delta^{\downarrow}_{\max}(\theta_{\mathrm{FPL}}; R)
  \;\ge\;
  R(1-\kappa)\,G_{\mathrm{base}}-LR^2.
  \label{eq:fpl-corrected-gain}
\end{equation}
Thus FPL strictly reduces this worst-case proxy drop whenever
\begin{equation}
 0<R<\frac{(1-\kappa)G_{\mathrm{base}}}{L}.
 \label{eq:fpl-local-radius}
\end{equation}
Here, the gradient-norm reduction $\kappa$ is given by Assumption~5, which the FPL objective is designed to induce.
\end{theorem}

\begin{proof}
By $L$-smoothness, throughout the stated neighborhoods,
\begin{equation}
 \left|L_{\mathrm{anti}}(\theta+\Delta)-L_{\mathrm{anti}}(\theta)
 -\nabla L_{\mathrm{anti}}(\theta)^\top\Delta\right|
 \le \frac{L}{2}\|\Delta\|^2.
\end{equation}
Consequently, for every $\|\Delta\|\le R$,
\begin{equation}
  L_{\mathrm{anti}}(\theta) - L_{\mathrm{anti}}(\theta+\Delta)
  \le
  -\,\nabla L_{\mathrm{anti}}(\theta)^\top \Delta
  +
  \frac{L}{2}\|\Delta\|^2.
\end{equation}
Maximizing over $\|\Delta\|\le R$ and using Cauchy--Schwarz yields
\begin{equation}
  \max_{\|\Delta\|\le R}
  \big\{
    L_{\mathrm{anti}}(\theta) - L_{\mathrm{anti}}(\theta+\Delta)
  \big\}
  \le
  R\,\big\|\nabla L_{\mathrm{anti}}(\theta)\big\|
  +
  \frac{L}{2}R^2.
\end{equation}
For $G(\theta)>0$, evaluate the other side of the smoothness inequality at $\Delta_*=-R\nabla L_{\mathrm{anti}}(\theta)/G(\theta)$:
\begin{equation}
 \Delta^{\downarrow}_{\max}(\theta;R)
 \ge L_{\mathrm{anti}}(\theta)-L_{\mathrm{anti}}(\theta+\Delta_*)
 \ge RG(\theta)-\frac{L}{2}R^2.
\end{equation}
If $G(\theta)=0$, the lower bound also holds since $\Delta=0$ gives a nonnegative maximum drop. Applying the lower bound to $\theta_{\mathrm{base}}$ and the upper bound to $\theta_{\mathrm{FPL}}$, then using Assumption~5, gives
\begin{align}
  \Delta^{\downarrow}_{\max}(\theta_{\mathrm{base}}; R)
  - \Delta^{\downarrow}_{\max}(\theta_{\mathrm{FPL}}; R)
  &\ge
  \Big(R\,G_{\mathrm{base}} - \frac{L}{2}R^2\Big)
  -
  \Big(R\,\kappa G_{\mathrm{base}} + \frac{L}{2}R^2\Big) \\
  &= R(1-\kappa)\,G_{\mathrm{base}}-LR^2.
\end{align}
The right-hand side is positive precisely under the stated local-radius condition.
\end{proof}

\noindent\textbf{Relation to experiments.} The guarantee applies where the $O(R)$ gradient benefit dominates the $O(R^2)$ curvature term. Empirically, the controlled FPL ablation keeps unmodified AUC at 0.99 while increasing post-merge AUC from 0.82 to 0.87 with $\beta:0\to5$ (PPL 4.18 to 4.22; Appendix~\ref{app:ablation-fpl}), and the merging and fine-tuning results in Section~\ref{sec:experiments} complement this local analysis on practical modifications.
\subsection{Computational Overhead of FPL}\label{app:TA_4}
In this section, we provide additional details on the computational overhead of the proposed Forgotten Perturbation-aware Learning (FPL). Although the optimization objective in Equation~\ref{eq:booster} involves second-order derivatives, we show that it can be solved efficiently without computing exact Hessian information.

\begin{equation}
 \arg \min_{\pi_\theta} \mathcal{L}_{\text{sim}}(\pi_\theta) + \beta \left( 
     \mathcal{L}_{\text{anti}}(\pi_\theta) - 
     \mathcal{L}_{\text{anti}}\!\left(\pi_\theta - \alpha \frac{\nabla \mathcal{L}_{\text{anti}}(\pi_\theta)}{\|\nabla \mathcal{L}_{\text{anti}}(\pi_\theta)\|}\right)
 \right)
\end{equation}

where $\mathcal{L}_{\text{anti}}(\cdot)$ denotes the forgetting loss on anti-watermarked texts. Intuitively, the second term measures the decrease of $\mathcal{L}_{\text{anti}}$ after one normalized fine-tuning step along the forgetting gradient. By minimizing this gap, the model not only learns a strong watermark but also remains robust to potential forgetting induced by user-side modifications.

To solve this perturbation minimization problem, we consider an iterative gradient method (e.g., SGD). By the chain rule, the update rule is:

\begin{equation}
\footnotesize
\pi_\theta^{t+1} = \pi_\theta^t - \eta \Bigg(
    \nabla \mathcal{L}_{\text{sim}}(\pi_\theta^t) + 
    \beta \Big(
        \nabla \mathcal{L}_{\text{anti}}(\pi_\theta^t) - 
        \nabla \mathcal{L}_{\text{anti}}\!\left(
            \pi_\theta^t - \alpha \frac{\nabla \mathcal{L}_{\text{anti}}(\pi_\theta^t)}{\|\nabla \mathcal{L}_{\text{anti}}(\pi_\theta^t)\|}
        \right)
        \underbrace{\nabla\!\left(\pi_\theta^t - \alpha \frac{\nabla \mathcal{L}_{\text{anti}}(\pi_\theta^t)}{\|\nabla \mathcal{L}_{\text{anti}}(\pi_\theta^t)\|}\right)}_{\text{second-order term}}
    \Big)
\Bigg)
\end{equation}

where $\eta$ is the learning rate. The last factor involves a second-order term (i.e., Hessian information), which is expensive to compute. Following prior work~\citep{finn2017model, rajeswaran2019meta}, we approximate this second-order term as a constant. The update rule then simplifies to:

\begin{equation}
\pi_\theta^{t+1} = \pi_\theta^t - \eta \Big(
    \nabla \mathcal{L}_{\text{sim}}(\pi_\theta^t) + 
    \beta \big(
        \nabla \mathcal{L}_{\text{anti}}(\pi_\theta^t) - 
        \nabla \mathcal{L}_{\text{anti}}\!\left(
            \pi_\theta^t - \alpha \frac{\nabla \mathcal{L}_{\text{anti}}(\pi_\theta^t)}{\|\nabla \mathcal{L}_{\text{anti}}(\pi_\theta^t)\|}
        \right)
    \big)
\Big)
\end{equation}

With this approximation, FPL requires only three forward/backward passes per optimization step:

\begin{enumerate}
    \item A forward pass to compute $\mathcal{L}_{\text{sim}}(\pi_\theta)$ on watermarked text.
    \item A forward and backward pass to compute $\nabla \mathcal{L}_{\text{anti}}(\pi_\theta)$ on anti-watermarked text.
    \item A forward pass to evaluate $\mathcal{L}_{\text{anti}}(\pi_\theta - \alpha \hat{g})$, where $\hat{g}$ is the normalized forgetting gradient.
\end{enumerate}

Thus, the overhead introduced by FPL is minimal. Moreover, CAWP accelerates convergence during training, making the overall computational cost comparable to baseline methods.

\section{Method Background}\label{sec:appendix-bg-method}
\subsection{Configurable Watermark Token Ratio $\gamma$}
\label{app:green_ratio}
In the main text, the watermark mapping model $M$ is trained to produce logits
that are approximately centered around zero, which results in an effective green
token ratio close to $0.5$. This choice is made for simplicity and does not
limit the flexibility of the framework. \ours{} also supports arbitrary watermark
token ratios $\gamma \in (0,1)$.

To target a desired ratio $\gamma$, we apply a simple scaling to the raw
watermark logits. Let $v_i = M(e_i) \in [-1,1]^{|\mathcal{V}|}$ denote the
watermark logits corresponding to the embedding $e_i$. We define a transformed
logit vector $s_i = [s_{i,j}]_{j=1}^{|\mathcal{V}|}$ as:
\[
s_{i,j} =
\begin{cases}
\frac{1-\gamma}{\gamma}\, v_{i,j}, & v_{i,j} > 0,\\[4pt]
v_{i,j}, & \text{otherwise}.
\end{cases}
\]
During training, the normalization objective is applied to $s_i$ in place of
$M(e_i)$, encouraging the scaled logits to realize the specified green token
proportion $\gamma$. The similarity objective $\mathcal{L}_{\text{sim}}$ remains
unchanged.

The ratio $\gamma$ specifies the token partition used during training. At detection time, \ours{} uses the same continuous policy score as in calibration, with the threshold estimated empirically on an independent null set for each configuration (Appendix~\ref{app:calibration}).

\subsection{Closed-source Watermark Schemes (KGW \& KTH)}\label{app:closed-source-bg}




{
\subsubsection{KGW}

The KGW watermark introduces a bias into the model’s decoding distribution\citep{kirchenbauer2023watermark}.  
Given the model output distribution $p$, the watermarked distribution is defined as
\begin{equation}
    f^{\text{KGW}}_w
    =
    \mathrm{softmax}\!\left(
        \log p
        + \delta \cdot
        h(x_{t-k+1:t}; \xi, \gamma, |\mathcal{V}|)
    \right),
\end{equation}
where $h(\cdot)$ produces a binary vector over the vocabulary that identifies the $\gamma$-portion of tokens to which the bias $\delta$ is applied. 
Its output is determined by a secret key $\xi$ and the previous $k$ tokens; when $k>1$, an additive hash is applied, and when $k=0$, a fixed green list is used\citep{zhao2024provable}.

\paragraph{Detection.}
Given a sequence $x$, the detector reconstructs the green list that would have been active at each step and checks whether the produced token belongs to it:
\begin{equation}
    S = 
    \sum_{t = k+1}^{\mathrm{len}(x)}
    h(x_{t-k:t-1})_{x_t}.
\end{equation}
For normal (unwatermarked) text, the matched tokens should appear only at the rate implied by the green-list proportion $\gamma$.  
If $S$ is noticeably larger than this expected level, it indicates that the generation process systematically preferred green-listed tokens, indicating the presence of the watermark.

}

{
\subsubsection{KTH}

The KTH watermark modifies decoding by selecting the token that best matches a key-dependent score.  
Given the model distribution $p$, the watermarked decoder is
\begin{equation}
\small
f^{\text{KTH}}_w(p, x, \xi)
=
\mathrm{onehot}\!\left(
\arg\max_i \frac{\xi^{(\mathrm{len}(x))}_i}{p_i},
|\mathcal{V}|
\right),
\end{equation}
where the key $\xi = (\xi^{(1)}, \ldots, \xi^{(m)})$ consists of $m$ vectors in $[0,1]^{|\mathcal{V}|}$, each sampled independently.  
This construction deterministically chooses the token with the largest key-to-probability ratio.

To increase variability across generations from the same prompt, the key is shifted by an offset $\tau$ before decoding, producing
\[
\xi' = (\xi^{(1+\tau \bmod m)}, \ldots, \xi^{(m+\tau \bmod m)}).
\]
A parameter $s \in [1,m]$ specifies the number of allowed shifts, taken as evenly spaced values
\[
\tau = \{\,i \cdot \lfloor m/s \rfloor \mid 0 \le i < s\,\}.
\]
Larger $s$ broadens the range of possible outputs and increases diversity.

\paragraph{Detection.}
To evaluate a candidate sequence $x$, the detector computes the alignment cost
\begin{equation}
\small
d(x, \xi)
=
\sum_{t=1}^{\mathrm{len}(x)}
\log\!\left( 1 - \xi^{(t)}_{x_t} \right),
\end{equation}
where lower values indicate that the text follows the key more closely and is therefore more likely watermarked.  
The significance of the observed score is assessed using a reference distribution obtained from non-watermarked text, and with a reference set of size $T$, the resulting $p$-value is bounded below by $1/(T+1)$.
}

\subsection{Model Modification Techniques}\label{app:modification-bg}
Open-source LLMs are subject to modifications for task adaptation or deployment efficiency.
The most prevalent types of such modifications include quantization, pruning, merging, and fine-tuning.


{
\subsubsection{Quantization}

Model quantization is an important technique for running large language models on memory-constrained hardware. It reduces the precision of parameters and activations from formats such as FP32 or BF16 to integer representations like INT8 or INT4, helping to lower storage use and speed up inference.

Quantization approaches can be grouped into weight-only quantization and calibration-guided quantization. Weight-only methods apply fixed quantizers directly to model weights without data calibration, such as LLM.int8() \citep{dettmers2022gpt3} and the NF4 format used in QLoRA \citep{dettmers2023qlora}. Calibration-guided methods optimize quantization parameters using a small calibration set to reduce reconstruction error. HQQ focuses on weight reconstruction, while GPTQ \citep{frantar2023optq} and AWQ \citep{lin2024awq} further reduce activation-level errors to improve the fidelity of the quantized model.

\subsubsection{Pruning}
Model pruning serves as a critical technique for optimizing large language models by systematically eliminating unnecessary parameters to achieve computational efficiency and reduced memory footprint. The fundamental principle involves identifying and removing weights that contribute minimally to model performance, thereby creating a more compact representation without substantial accuracy degradation.

Pruning strategies divide into two primary paradigms: unstructured and structured methodologies. In unstructured pruning, individual parameters are evaluated and removed based on their importance, creating irregular sparsity patterns throughout the network. This granular approach enables precise control over sparsity levels but often requires specialized hardware for efficient inference. Conversely, structured pruning eliminates coherent parameter groups—including entire channels, filters, attention mechanisms, or even complete layers—resulting in regular sparsity patterns that align naturally with existing hardware architectures.

The optimization objective in pruning centers on preserving the functional behavior of the original model $f(\cdot; \boldsymbol{\theta})$ after parameter removal. Given a calibration dataset $\mathcal{D}$, the pruning process seeks to identify sparse parameters $\boldsymbol{\theta}'$ that minimize the divergence between the original and compressed models:
\begin{equation}
\min_{\boldsymbol{\theta}'} \sum_{x \in \mathcal{D}} \| f(x; \boldsymbol{\theta}) - f(x; \boldsymbol{\theta}') \|^2,
\end{equation}
Here, $\boldsymbol{\theta}'$ represents the parameter set after pruning, where numerous weights have been eliminated through zeroing (unstructured) or complete removal (structured).

Contemporary unstructured approaches like \textsc{Wanda} \citep{sun2024a}, \textsc{SparseGPT} \citep{frantar2023sparsegpt}, and \textsc{GBLM} \citep{das2023beyond} employ sophisticated scoring mechanisms to assess parameter significance at the individual weight level, enabling high sparsity ratios while maintaining model quality. Meanwhile, structured techniques including \textsc{Sheared LLaMA} \citep{xia2023sheared} and \textsc{LLM-Pruner} \citep{ma2023llm} adopt a holistic approach, jointly optimizing the removal of architectural components to achieve efficient compression while adhering to the reconstruction objective.

\subsubsection{Model Merging}
Model merging represents an innovative paradigm for synthesizing multiple specialized models into a unified architecture by strategically combining their learned parameters. This approach enables the integration of diverse capabilities acquired through task-specific training, effectively consolidating expertise from different domains into a single deployable model without requiring additional training.

Recent investigations \citep{yang2024model, goddard2024arcee} have revealed that parameter-space merging can successfully preserve and even amplify performance across multiple tasks, suggesting that distinct model capabilities can coexist within a shared parameter framework. The theoretical foundation rests on the notion of \emph{task vectors}—the hypothesis that different model competencies manifest as approximately orthogonal directions within the high-dimensional parameter space, enabling their combination through various mathematical operations including weighted averaging and geometric interpolation.

Among the various merging strategies, \emph{Spherical Linear Interpolation} (\textsc{SLERP}) emerges as a mathematically rigorous approach that respects the geometric structure of the parameter space. Unlike naive linear interpolation, \textsc{SLERP} traces the geodesic path between model parameters on the surface of a hypersphere, maintaining constant norm throughout the interpolation process. For two parameter configurations $\boldsymbol{\theta}_1$ and $\boldsymbol{\theta}_2$ separated by angle $\Omega$, the interpolated parameters at position $t \in [0,1]$ are computed as:
\begin{equation}
\text{SLERP}(\boldsymbol{\theta}_1, \boldsymbol{\theta}_2, t) = 
\frac{\sin[(1 - t)\Omega]}{\sin \Omega} \boldsymbol{\theta}_1 + 
\frac{\sin[t\Omega]}{\sin \Omega} \boldsymbol{\theta}_2.
\end{equation}
This spherical interpolation maintains the magnitude relationships inherent in the original models, potentially preserving learned representations more faithfully than alternative methods. When the parameter vectors align perfectly ($\Omega = 0$), the formulation gracefully degenerates to standard linear interpolation, ensuring computational stability across all configurations. The geometric preservation properties of \textsc{SLERP} make it particularly suitable for merging models where maintaining the relative structure of learned features is crucial for performance retention.

\subsubsection{Fine-tuning}
Model finetuning \citep{zhang2023instruction, zhang2024scaling} constitutes a fundamental adaptation methodology that transforms general-purpose language models into specialized systems through targeted training on domain-specific or task-oriented datasets. This technique bridges the gap between broad pretraining objectives and precise downstream requirements, enabling models to acquire specialized knowledge and behaviors absent from their initial training distribution.

The finetuning paradigm encompasses several distinct approaches, notably \emph{Supervised Finetuning (SFT)} and \emph{Instruction Tuning}. SFT directly optimizes model parameters using labeled examples from the target domain, establishing direct mappings between inputs and desired outputs. Instruction tuning extends this framework by training models to interpret and execute natural language directives across heterogeneous tasks, fostering more flexible and generalizable behavior. This instruction-following capability proves essential for developing conversational AI systems that can engage in open-ended dialogue while maintaining alignment with user intentions.

The optimization framework underlying both approaches seeks to minimize prediction error on the adaptation dataset. Starting from pretrained parameters $\boldsymbol{\theta}$, the finetuning process identifies updated parameters $\boldsymbol{\theta}'$ that optimize performance on the target dataset $\mathcal{D} = \{(x_i, y_i)\}_{i=1}^N$:
\begin{equation}
\min_{\boldsymbol{\theta}'} \frac{1}{N} \sum_{i=1}^{N} \mathcal{L}(f(x_i; \boldsymbol{\theta}'), y_i),
\end{equation}
where $\mathcal{L}$ denotes the task-specific loss function measuring the discrepancy between model predictions and ground truth labels.

Contemporary finetuning strategies range from full-model updates to parameter-efficient alternatives. While comprehensive finetuning modifies all model parameters, resource-constrained scenarios benefit from selective adaptation techniques like LoRA (Low-Rank Adaptation) or adapter modules. These methods introduce minimal trainable parameters while freezing the majority of the pretrained model, achieving computational efficiency without compromising adaptation quality. Such parameter-efficient approaches have become increasingly vital as model scales continue to expand, enabling practical deployment of specialized models across diverse computational environments.
}

\section{Experimental Configuration}\label{sec:appendix-config}
\subsection{Baseline Setting}
\label{Appendix:baseline}
 Each table lists the hyperparameter configurations for the four open-source watermarking methods. The columns Left, Middle, and Right indicate the parameter settings used for the leftmost, middle, and rightmost points on each method’s curve in the plots. Every \ours{} entry uses $n=10$ words on all three models; only $\delta$ varies across the \ours{} points.
 
\begin{table}[t]
\centering
\small
\caption{Watermarking hyperparameter configurations for LLaMA-8B.}
\setlength{\tabcolsep}{6pt} 
\begin{tabular}{@{}lccc@{}}
\toprule
\textbf{Method} & \textbf{Left} & \textbf{Middle} & \textbf{Right} \\
\midrule
Gu et al. (KGW)       & $n=0$, $\delta=1$         & $n=1$, $\delta=1$         & $n=1$, $\delta=2$         \\
Gu et al. (KTH)       & $s=4$                    & $s=2$                    & $s=1$                    \\
Gloaguen et al. (KGW) & $n=0$, $\delta=1$ & 
$n=1$, $\delta=1$ & 
$n=1$, $\delta=2$ \\
\ours{}                    &$\delta=0.3$         & $\delta=0.5$       & $\delta=1$         \\
\bottomrule
\end{tabular}
\label{tab:watermark-config-llama8b}
\end{table}

\begin{table}[t]
\small
\centering
\caption{Watermarking hyperparameter configurations for LLaMA-3.2-3B.}
\begin{tabular}{@{}lccc@{}}
\toprule
\textbf{Method} & \textbf{Left} & \textbf{Middle} & \textbf{Right} \\
\midrule
Gu et al. (KGW)       & $n=0$, $\delta=1$         & $n=1$, $\delta=1$         & $n=1$, $\delta=2$         \\
Gu et al. (KTH)       & $s=4$                    & $s=2$                    & $s=1$                    \\
Gloaguen et al. (KGW) & $n=0$, $\delta=1$ & 
$n=1$, $\delta=1$ & 
$n=1$, $\delta=2$ \\
\ours{}                    &$\delta=0.5$         & $\delta=1$       & $\delta=1.5$         \\
\bottomrule
\end{tabular}
\label{tab:watermark-config-llama32}
\end{table}

\begin{table}[t]
\small
\centering
\setlength{\tabcolsep}{6pt} 
\caption{Watermarking hyperparameter configurations for Phi-2-2.7B.}
\begin{tabular}{@{}lccc@{}}
\toprule
\textbf{Method} & \textbf{Left} & \textbf{Middle} & \textbf{Right} \\
\midrule
Gu et al. (KGW)       & $n=0$, $\delta=1$         & $n=1$, $\delta=1$         & $n=1$, $\delta=2$         \\
Gu et al. (KTH)       & $s=4$                    & $s=2$                    & $s=1$                    \\
Gloaguen et al. (KGW) & $n=0$, $\delta=1$ & 
$n=1$, $\delta=1$ & 
$n=1$, $\delta=2$ \\
\ours{}                    &$\delta=0.5$         & $\delta=1$       & $\delta=1.5$         \\
\bottomrule
\end{tabular}

\label{tab:watermark-config-phi2}
\end{table}

\subsection{Training Configurations}\label{app:training-config}
\label{appendix:training}
In our watermarking pipeline, the semantic embeddings are generated by \texttt{compositional-bert-large}~\citep{chanchani2023composition}. During training, both the student LLM $\pi_\theta$ and the watermark mapping model $M$ are jointly optimized to maximize learnability and detection consistency. We set the regularization weights $\lambda_1 = \lambda_2 = 1$ in the training loss for CAWP. For the FPL component, we use perturbation step size $\alpha = 0.1$ and regularization weight $\beta = 5$. In our ablation study, we further analyze the sensitivity of watermark robustness to variations in these hyperparameters.

All models are fine-tuned on subsets of OpenWebText with different watermark strategies using a batch size of 64 sequences, sequence length of 512 tokens, a maximum learning rate of 1e-5 with cosine decay, and a linear warmup over the first 10\% of steps. We use the AdamW optimizer with $(\beta_1, \beta_2) = (0.9, 0.999)$ and no weight decay. For KTH watermark distillation, we follow the same setup except for using a batch size of 128 and a sequence length of 256 tokens to accommodate memory constraints. For the Gloaguen et al.~\citep{gloaguen2025towards}-KGW baseline, we follow their configuration. Starting from a model distilled with KGW, we fine-tune it on OpenWebText for 2{,}500 steps with a batch size of 64, sequence length of 512 tokens, a learning rate of 1e-5, and the AdamW optimizer. A cosine learning rate schedule is applied with 500 warmup steps. To identify stable parameters, we compute contrastive task vectors based on parameter change before and after fine-tuning, and perform a second-stage distillation restricted to these stable weights. Each training run took approximately 6 hours on 1 NVIDIA B200 192GB GPUs.

\subsection{Model Modification Settings}
\label{appendix:modification}
To assess robustness, we apply four types of model modifications that simulate real-world user behavior: (1) quantization, including INT8~\citep{dettmers2022gpt3} and GPTQ~\citep{frantar2023optq} at 8-bit precision, GPTQ\citep{frantar2023optq} and HQQ~\citep{badri2023hqq} at 4-bit; (2) unstructured pruning using WANDA~\citep{sun2024a} and SparseGPT ~\citep{frantar2023sparsegpt} at 20\% and 50\% sparsity levels; (3) model merging via SLERP~\citep{goddard2024arcee}, where the watermarked model is interpolated with its non-watermarked base model using mixing ratios from 0.1 to 0.9 following ~\citep{gloaguen2025towards}; and (4) full-parameter fine-tuning on the task-specific \texttt{OPENMATHINSTRUCT} dataset \citep{toshniwal2024openmathinstruct} and \texttt{OPENCODEINSTRUCT} dataset \citep{ahmad2025opencodeinstruct}, reflecting the common use case where LLM users fine-tune open-source models on downstream data to build domain-specific experts. All modifications are implemented following the original settings of their respective methods.

\section{Additional Empirical Results}\label{sec:appendix-results}

\subsection{Computational Efficiency Analysis}\label{app:efficiency}
We provide a detailed comparison of training cost with prior watermarking methods. Our proposed techniques do not incur significant overhead and, in fact, accelerate convergence.

\noindent\textbf{CAWP.} The first component, CAWP, introduces only a lightweight trainable MLP as the watermark mapping model $M$, which contains $1.16$M parameters, just $0.0388\%$ of a $3$B-parameter LLM. This addition is negligible compared to the base LLM and does not substantially increase computational cost.

\noindent\textbf{FPL.} The second component, FPL, involves two additional forward and backward passes per iteration. However, by using a learning-friendly watermark, our method reduces the number of training steps needed to achieve high performance. For example, \ours{} reaches an AUC of $0.997$ within $2000$ steps, while KGW requires $5000$ steps to reach a maximum AUC of $0.991$.

\noindent\textbf{Wall-clock Training Time.} In practice, the total wall-clock training time remains comparable across methods. For 8B-parameter models trained on $1\times$B200 192GB GPUs, all methods (KGW, KTH, \ours{}) complete training in about $6$ hours. Thus, despite the slightly higher per-iteration cost, \ours{} achieves better performance with no additional end-to-end training time.

\subsection{Cross-Distribution and Cross-Architecture Generalization}\label{app:cross-arch}
\begin{figure}[h]
\centering
\includegraphics[width=0.6\linewidth]{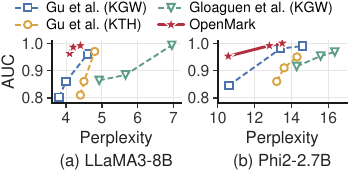}
\caption{\ours{} trained on C4 with LLaMA-3-8B, evaluated on Wikipedia prompts without retraining.}
\label{fig:generalization}
\end{figure}

\noindent\textbf{OOD prompts.} A practical deployment concern is whether \ours{} remains effective under prompts that differ from the C4 training distribution. We evaluate the C4-trained \ours{} model on Wikipedia prompts using the Figure~\ref{fig:TPR-FPR-PPL} pipeline, without retraining the student LLM. Figure~\ref{fig:generalization} shows that \ours{} preserves the best detectability-quality trade-off under this distribution shift, suggesting that the learned watermark policy is not tied to C4.

\noindent\textbf{Other architectures.} We further evaluate whether \ours{} generalizes beyond LLaMA-family models. Applying our pipeline to GPT-J-6B (a non-LLaMA architecture) using the same settings as Figure~\ref{fig:TPR-FPR-PPL}, \ours{} consistently outperforms baselines: at AUC$>0.99$, \ours{} maintains a perplexity of $17.8$, while existing methods require at least $21.5$.

\begin{wrapfigure}{r}{0.4\textwidth}
    \centering
    \includegraphics[width=\linewidth]{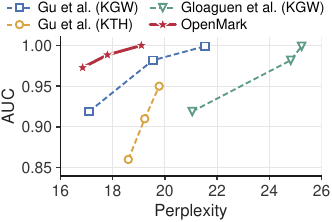}
    \caption{\ours{} results on GPT-J-6B, demonstrating generalization beyond LLaMA-family architectures.}
        \vspace{-10pt}
    \label{fig:GPT-J}
\end{wrapfigure}

\subsection{Ablation on FPL Hyperparameters}\label{app:ablation-fpl}
We conduct an ablation study by varying the perturbation step size $\alpha$ (Table~\ref{tab:ablation-alpha}) and the regularization weight $\beta$ (Table~\ref{tab:ablation-beta}) in Equation~\ref{eq:booster}. When both are set to zero, the model reduces to using CAWP alone. In this case, robustness improves slightly over the baseline (Merge $\text{AUC}=0.82$ after merging), indicating CAWP alone can already enhance watermark robustness. Introducing FPL via non-zero $\beta$ brings modest but consistent gains in robustness ($\text{AUC}=0.87$ at $\alpha = 0.1$, $\beta = 5$), though with a mild increase in perplexity. {However, larger values of $\alpha$ or $\beta$ lead to degraded or unstable results, highlighting the importance of careful hyperparameter tuning to balance robustness and generation quality.}

\begin{table}[h]
\centering
\small
\caption{Ablations on perturbation step size $\alpha$ and regularization weight $\beta$. We report AUC and PPL of the unaltered model and AUC after merging ($t = 0.5$) to indicate robustness.}
\begin{subtable}[t]{0.52\textwidth}
\centering
\caption{Perturbation step size $\alpha$.}
\label{tab:ablation-alpha}
\setlength{\tabcolsep}{5pt}
\begin{tabular}{lccccc}
\toprule
$\boldsymbol{\alpha}$ & 0 & \textbf{0.1} & 1 & 2 & 5 \\
\midrule
AUC         & $0.99$ & \bm{$0.99$} & $0.98$ & $0.97$ & $0.95$ \\
PPL         & $4.18$ & \bm{$4.22$} & $4.30$ & $4.70$ & $5.00$ \\
Merge AUC   & $0.82$ & \bm{$0.87$} & $0.86$ & $0.83$ & $0.79$ \\
\bottomrule
\end{tabular}
\end{subtable}
\hfill
\begin{subtable}[t]{0.44\textwidth}
\centering
\caption{Regularization weight $\beta$.}
\label{tab:ablation-beta}
\setlength{\tabcolsep}{5pt}
\begin{tabular}{lcccc}
\toprule
$\boldsymbol{\beta}$ & 0 & 1 & \textbf{5} & 10 \\
\midrule
AUC         & $0.99$ & $1.00$ & \bm{$0.99$} & $0.95$ \\
PPL         & $4.18$ & $4.21$ & \bm{$4.22$} & $4.52$ \\
Merge AUC   & $0.82$ & $0.85$ & \bm{$0.87$} & $0.84$ \\
\bottomrule
\end{tabular}
\end{subtable}
\end{table}

\subsection{Visualization of ROC Curves}\label{app:ROC}
\begin{wrapfigure}{r}{0.3\textwidth}
\vspace{-0.1in}
\centering
\includegraphics[width=\linewidth]{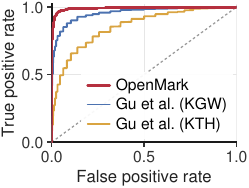}
\caption{ROC curves for different methods.}
\vspace{-0.2in}
\label{fig:roc}
\end{wrapfigure}

In Figure~\ref{fig:roc}, we visualize the ROC curves for the proposed \ours{} and the state-of-the-art method by Gu et al. We select watermarked LLaMA3-8B models with the same level of perplexity (i.e., $4.7$). The results indicate that \ours{} achieves better watermark detectability while maintaining the same level of generation quality. 

Specifically, the ROC curve of \ours{} closely follows the top-left corner, indicating a higher true positive rate (TPR) across nearly all false positive rate (FPR) thresholds. In contrast, the methods by Gu et al. show noticeably lower TPRs, particularly at low FPR regions, which suggests a less reliable distinction between watermarked and non-watermarked texts under tight detection constraints. This improvement is especially evident in the early phase of the curve (e.g., $\text{FPR} < 0.1$), where \ours{} already achieves near-perfect detection performance, while the baseline methods lag behind. These findings further confirm that \ours{} not only embeds robust watermark patterns but also enables more confident detection, even at low error tolerance levels, making it more suitable for security-critical applications.

\subsection{Detection Performance under Different FPRs}
\label{app:fpr_tpr}

We further evaluate detection robustness by measuring the true positive rate (TPR) at different false positive rate (FPR) levels. 
This complements the AUC metric and highlights performance in stricter detection regimes. 
As shown in Table~\ref{tab:fpr_tpr}, \ours{} consistently outperforms prior methods, especially under very low FPRs (e.g., 0.1\%), which are critical for practical deployment. 

\begin{table}[h]
\centering\small
\caption{Comparison of detection performance (TPR at different FPR levels) on LLaMA-3-8B. \ours{} demonstrates significantly stronger detection under stringent low-FPR conditions.}
\label{tab:fpr_tpr}
\setlength{\tabcolsep}{10pt}
\renewcommand{\arraystretch}{0.95}
\begin{tabular}{lccc}
\toprule
Method & TPR@0.1\% $\uparrow$ & TPR@1\% $\uparrow$ & TPR@10\% $\uparrow$ \\
\midrule
Gu et al. (KTH) & 25.8\% & 53.6\% & 84.0\% \\
Gu et al. (KGW) & 54.2\% & 82.7\% & 97.5\% \\
\ours{} & \textbf{78.1\%} & \textbf{92.3\%} & \textbf{99.5\%} \\
\bottomrule
\end{tabular}
\end{table}


\subsection{Overall Detection AUC}\label{app:AUC}
\begin{figure}[h!]
    \centering
    \includegraphics[width=\linewidth]{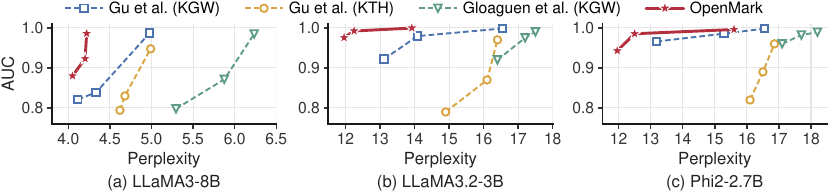}
    \caption{Effectiveness comparison of different open-source LLMs text watermarks, in terms of detection AUC and median PPL on three LLMs. Better watermark detectability and generation quality are indicated by higher AUC and lower PPL, as shown by lines closer to the upper left corner.}
    \label{fig:AUC-PPL}
\end{figure}

We compared our \ours{} and baselines in terms of detection AUC and perplexity in Figure~\ref{fig:AUC-PPL}, the results show that our method \ours{} can achieve the best AUC while maintaining the lowest perplexity.

\subsection{Robustness against Paraphrasing Attack}
\label{app:paraphrasing_attack}
We evaluate the robustness of our watermarking method under paraphrasing using
DIPPER\citep{krishna2023paraphrasing}, a controllable paraphraser that rewrites text while preserving semantics in Table~\ref{tab:dipper}.
We use two settings: DIPPER-1, with lexical diversity set to $60$ and order diversity $0$; and DIPPER-2, with lexical diversity $60$ and order diversity $20$. \ours{} achieves AUC 0.90 and 0.87 under the two settings, versus 0.86 and 0.84 for the strongest baseline, Gu et al.-KGW. We further report TPR at 1\% FPR under stronger rewriting attacks below.
\input{tables/rewrite}

\paragraph{Stronger rewriting attacks.}
We evaluate three attacks on 200-token generations from the released LLaMA3-8B model, using each method's human-C4-calibrated threshold at nominal 1\% FPR. For independent paraphrasing, Qwen3.5-9B rewrites each passage once with an instruction to preserve meaning. Repeated paraphrasing applies the same model for three consecutive rounds. For watermark-aware rewriting, the attacker receives each method's private watermark artifact (\ours{}'s detector or KGW's hash key and tokenizer), generates eight diverse candidates, and selects the lowest-scoring one. This oracle attack grants stronger access than the deployment threat model.

\begin{table}[ht]
\centering\small
\caption{TPR at nominal 1\% FPR under stronger text rewriting. The final row is a detector-oracle stress test.}
\label{tab:strong-rewriting}
\begin{tabular}{lcc}
\toprule
Attack & Gu et al.-KGW & \ours{} \\
\midrule
Independent paraphrasing & 0.25 & 0.32 \\
Repeated paraphrasing, three rounds & 0.12 & 0.17 \\
Watermark-aware rewriting, eight candidates & 0.00 & 0.03 \\
\bottomrule
\end{tabular}
\end{table}

\ours{} retains higher TPR than Gu et al.-KGW under all three attacks, including the detector-oracle setting that exceeds our threat model.
\subsection{Robustness against Model Modifications across Downstream Tasks}\label{app:modifications_tasks}
We conducted experiments to evaluate FPL's effectiveness on more diverse downstream tasks. Specifically, we fine-tuned on the Alpaca dataset to further examine watermark robustness beyond code and math domains. As shown in Table~\ref{tab:ft_steps}, our method consistently outperforms the best baseline (\citeauthor{gloaguen2025towards} with KGW), with \ours{}'s relative improvements shown in parentheses.
At 1\% FPR, \ours{}'s advantage over Gloaguen et al.-KGW grows from +2.8\% after 300 steps to +16.5\% after 1{,}500 steps (TPR 54.3\% and 16.9\%), indicating slower decay under longer fine-tuning.

\begin{table}[h]
\centering\small
\caption{Performance of watermarked LLaMA-3-8B under different fine-tuning steps $s$. 
The values in parentheses (prefixed with $+$) indicate the relative improvement compared to the baseline.}
\label{tab:ft_steps}
\begin{tabular}{cccc}
\toprule
Step $s$ & AUC & TPR@1\% & TPR@10\% \\\midrule
$300$  & $0.91$ \improve{(+0.02)} & $54.3\%$ \improve{(+2.8\%)} & $72.5\%$ \improve{(+6.0\%)} \\
$600$  & $0.87$ \improve{(+0.04)} & $49.4\%$ \improve{(+4.0\%)} & $69.9\%$ \improve{(+2.9\%)} \\
$900$  & $0.85$ \improve{(+0.02)} & $40.5\%$ \improve{(+9.1\%)} & $60.2\%$ \improve{(+6.9\%)} \\
$1200$ & $0.81$ \improve{(+0.04)} & $26.8\%$ \improve{(+13.0\%)} & $44.7\%$ \improve{(+13.6\%)} \\
$1500$ & $0.79$ \improve{(+0.05)} & $16.9\%$ \improve{(+16.5\%)} & $33.3\%$ \improve{(+21.0\%)} \\\bottomrule
\end{tabular}
\end{table}

\subsection{Compared with Classifier-based LLM text Detectors}\label{app:detectors}
One distinct advantage of watermarking over classifier-based detectors is its ability to attribute text to a specific model, rather than just distinguishing LLM and human text. We evaluated \ours{} using the open-source classifier SuperAnnotate/ai-detector, which flagged 90.6\% of \ours{}-generated texts as AI-generated. It also flagged ~90\% of outputs from non-watermarked and differently watermarked LLMs (Qwen-4B/8B, GPT-6B, LLaMA3-3B) as LLM-generated. To further analyze this, we fine-tuned a binary RoBERTa classifier on texts from LLaMA3-8B and human texts. It detected 60–80\% of outputs from other LLMs (Qwen-4B/8B, GPT-6B, LLaMA3-3B) as LLaMA3-8B-generated. This shows that classifier-based methods primarily learn to distinguish LLM vs. human text, not between different LLMs. In contrast, our \ours{} watermark enables model-level attribution. Specifically, we used the watermark policy model trained with LLaMA3-8B (student LLM) to detect outputs from other LLMs. The false positive rates remained low: 0.8\% (Qwen-4B), 1.0\% (Qwen-8B), 1.1\% (GPT-6B), and 0.6\% (LLaMA3-3B). This highlights \ours{}’s precision in determining whether a text was generated by a specific model.

\subsection{Cross-Model Attribution across \ours{}-Watermarked Models}\label{app:cross-attribution}
We evaluate each detector $M_i$ on outputs of all three independently watermarked generators $G_j$. A matched pair ($i=j$) measures detection of the model's own outputs. A mismatched pair ($i\ne j$) measures how often the detector flags another watermarked model's outputs. Section~\ref{sec:results}, Table~\ref{tab:cross-attribution}, reports all nine rates.

\subsection{Different Embedding Model and Watermark Mapping Model}\label{app:embedding_mapping}

To further validate the generality of our method, we supplemented the experiments with two additional embedding models: \texttt{thenlper/gte-large} and \texttt{intfloat/e5-large-v2}. Results in Table~\ref{tab:embedding} show that our method maintains strong detection performance across different embeddings, consistently achieving near-perfect AUC and high TPR scores.

\begin{table}[h]
\centering
\caption{Performance of \ours{} watermarking with LLaMA-3.2-3B under different embedding models.}
\label{tab:embedding}
\setlength{\tabcolsep}{6pt}
\renewcommand{\arraystretch}{0.95}
\begin{tabular}{lccc}
\toprule
Embedding Model & AUC & TPR@1\% & TPR@10\% \\
\midrule
\texttt{thenlper/gte-large}   & $0.997$ & $95.2\%$ & $99.2\%$ \\
\texttt{intfloat/e5-large-v2} & $0.994$ & $94.7\%$ & $99.0\%$ \\
\bottomrule
\end{tabular}
\end{table}

We also compared different architectures for the watermark mapping model $M$. An MLP yields the best performance, as it can effectively exploit the full semantic embedding. By contrast, convolutional neural networks (CNNs) perform significantly worse, since semantic embeddings are single dense vectors without spatial or sequential structure for convolution to leverage. Thus, we adopt the MLP design in our framework.

\subsection{Distillation Data Scaling Study}\label{app:distillation}
Recent work by \cite{gloaguen2025towards} shows that watermark robustness can be improved by substantially enlarging the distillation dataset~\cite{li2025gvd}. In Table~\ref{tab:robustness}, \ours{} uses a 64 million tokens distillation corpus to maintain the trade-off between watermark robustness and model fidelity. To examine how \ours{} benefits from data scaling, we conduct additional experiments where the distillation corpus is expanded from 32M to 160M tokens. 
Table~\ref{tab:data-scaling} summarizes the results under two robustness settings: (1) model merging (ratio=0.5) and (2) 1500-step fine-tuning on OpenMathInstruct~\citep{toshniwal2024openmathinstruct}. We observe consistent improvements in TPR@5\% as the distillation data increases. For example, TPR@5\% FPR improves from 0.49 to 0.69 under merge-0.5, and from 0.29 to 0.72 under fine-tuning. The perplexity increases only moderately (4.0 to 5.3), indicating that data scaling does not substantially harm model quality.

\input{tables/scaling}

\input{sections/camera_ready_results}

\FloatBarrier  
\section{Limitations}\label{app:limitations}
\paragraph{Platform and agent scope.} \oursx{} provides built-in model workflows and extension interfaces, but support depends on the chosen adapter, watermark method, and evaluation stage. Additional integration is required for arbitrary new adapters. The current agent workflow detects generated final answers, not complete action trajectories; the multimodal demonstrations use compatible methods rather than the full \ours{} procedure in every modality. These demonstrations do not establish robustness for every method--model--modification combination.
\paragraph{Provenance and forgery.} \ours{} targets model-level provenance; user-level attribution and cryptographically unforgeable watermarking~\citep{liu2023unforgeable} are complementary directions.

\paragraph{Private artifacts and extraction.} Our deployment assumes the teacher $\pi_o$ and the policy $M$ remain private; formal guarantees against policy extraction are left to future work.

\paragraph{Bounded robustness.} Since a user can arbitrarily overwrite the released weights, no weight-based watermark can survive unbounded modification. \ours{} slows degradation under practical modifications, but detection still decreases under heavier merging, longer fine-tuning, and strong rewriting (Section~\ref{sec:experiments}, Appendix~\ref{app:paraphrasing_attack}). Our theoretical robustness result is local and stated for the anti-watermark proxy loss.

\section{Impact Statement}\label{app:impact}

This work advances generation watermarking through the \oursx{} platform and the \ours{} method. The platform supports configurable evaluation of open model and agent outputs, while the method improves text-watermark detectability and robustness under the reported protocols. The modality-specific workflows broaden the kinds of generated outputs that can be studied; detection remains statistical evidence under a specified key and protocol.
We do not identify any direct or immediate ethical concerns specific to this work.



%% file: tables/rewrite.tex
\begin{table}[h]
\centering
\caption{AUC under DIPPER attacks.}
\label{tab:dipper}
\small
\begin{tabular}{lcc}
\toprule
Watermark Method & DIPPER1  & DIPPER2 \\
\midrule
\citet{gu2024on}-KTH & $0.82$ & $0.79$ \\
\citet{gu2024on}-KGW & $0.86$ & $0.84$ \\
\citet{gloaguen2025towards}-KGW & $0.80$ & $0.74$ \\
\textbf{\ours{} (Ours)} & \bm{$0.90$} & \bm{$0.87$} \\
\bottomrule
\end{tabular}
\end{table}

%% file: tables/scaling.tex
\begin{table}[h!]
\centering
\small
\caption{Effect of distillation data scaling on watermark robustness. Larger distillation datasets improve TPR@5\%FPR under both merge-0.5 and FT-1500 settings, with only moderate PPL increase.}
\begin{tabular}{lccccc}
\toprule
\textbf{Modification} & \textbf{32M} & \textbf{64M} & \textbf{96M} & \textbf{128M} & \textbf{160M} \\
\midrule
Merge-0.5 & 0.49 & 0.57 & 0.63 & 0.67 & 0.69 \\
FT-1500 & 0.29 & 0.37 & 0.50 & 0.63 & 0.72 \\
PPL & 4.0 & 4.2 & 4.7 & 5.1 & 5.3 \\
\bottomrule
\end{tabular}
\label{tab:data-scaling}
\end{table}

%% file: sections/camera_ready_results.tex
\subsection{Null Calibration and Text Length}\label{app:calibration}
The main null consists of 5{,}000 held-out human-C4 continuations. Calibration uses a separate, non-overlapping set of 200-token human-C4 texts. For each \ours{} detector, we set $\tau_{\mathrm{C4}}$ to the empirical 99th-percentile calibration score and freeze it before evaluation. The same threshold is applied to all null sources below without recalibration.

\begin{table}[ht]
\centering\small
\caption{Realized FPR on disjoint 200-token null sources at the same frozen human-C4 threshold. \ours{}'s watermarked outputs have TPR 91.8\% at this threshold.}
\label{tab:null-transfer}
\begin{tabular}{lc}
\toprule
Evaluation null & Realized FPR \\
\midrule
Held-out human C4 & 1.06\% \\
Unwatermarked same-base generations & 1.28\% \\
Qwen3-4B generations & 0.82\% \\
\bottomrule
\end{tabular}
\end{table}

The realized FPRs stay close to the nominal 1\% across all three null sources, so the human-C4 threshold transfers without source-specific recalibration. With the 200-token threshold frozen, longer texts yield lower FPR and higher TPR (Table~\ref{tab:length-calibration}), since they provide more watermark evidence and lower null-score variance; length-specific calibration can target the nominal FPR at each length.

\begin{table}[ht]
\centering\small
\caption{Length transfer of a single detector and threshold calibrated on 200-token human-C4 text.}
\label{tab:length-calibration}
\begin{tabular}{lcccc}
\toprule
Text length (tokens) & 100 & 200 & 300 & 500 \\
\midrule
Realized FPR & 1.42\% & 1.06\% & 0.72\% & 0.38\% \\
TPR & 86.3\% & 91.8\% & 94.1\% & 96.6\% \\
\bottomrule
\end{tabular}
\end{table}

For shorter texts, we rescore 200 watermarked outputs of \ours{} trained on Granite-4.2-3B, together with 500 calibration and 500 test human-C4 texts, each truncated to its first $L$ tokens and scored without the prompt. The threshold is calibrated at each length for 1\% FPR (Table~\ref{tab:short-text}). Because the first ten words serve as context, a text of $L$ tokens has about $L-12$ scored tokens. AUC is 0.977 at 50 tokens and 0.997 at 100 tokens. At the per-length threshold, TPR reaches 93.5\% at 100 tokens and 71.5\% at 50 tokens, while the realized FPR stays at or below 1.2\%; the 200-token threshold, in contrast, admits 8.2\% and 22.4\% FPR at 50 and 25 tokens. A 10-token text has fewer than ten words and thus no scored token; scoring with a 3-word context instead gives AUC 0.739 and TPR 9.5\% at 10 tokens, and AUC 0.877 and TPR 28.5\% at 20 tokens.

\begin{table}[ht]
\centering\small
\caption{Short-text detection with \ours{} on Granite-4.2-3B, scored on the text alone with a threshold calibrated at each length.}
\label{tab:short-text}
\begin{tabular}{lcccc}
\toprule
Text length (tokens) & 25 & 50 & 100 & 200 \\
\midrule
Scored tokens (mean) & 12.7 & 37.7 & 87.7 & 187.7 \\
AUC & 0.888 & 0.977 & 0.997 & 1.000 \\
TPR & 30.0\% & 71.5\% & 93.5\% & 100\% \\
Realized FPR & 1.2\% & 0.6\% & 0.2\% & 0.2\% \\
FPR at the 200-token threshold & 22.4\% & 8.2\% & 1.4\% & 0.2\% \\
\bottomrule
\end{tabular}
\end{table}

Table~\ref{tab:short-text-main} compares the same length-specific detection task with Gu et al.-KGW. The 10-token row uses a 3-word context.

\begin{table}[ht]
\centering\small
\caption{Detection on short texts from Granite-4.2-3B. TPR is measured at 1\% FPR with a threshold calibrated for each length.}
\label{tab:short-text-main}
\begin{tabular}{ccccc}
\toprule
\multirow{2}{*}{Tokens} & \multicolumn{2}{c}{Gu et al.-KGW} & \multicolumn{2}{c}{\ours{}} \\
\cmidrule(lr){2-3} \cmidrule(lr){4-5}
& AUC & TPR@1\% & AUC & TPR@1\% \\
\midrule
10  & 0.667 & 0.015 & 0.739 & 0.095 \\
25  & 0.758 & 0.205 & 0.888 & 0.300 \\
50  & 0.936 & 0.465 & 0.977 & 0.715 \\
100 & 0.971 & 0.775 & 0.997 & 0.935 \\
200 & 0.983 & 0.870 & 1.000 & 1.000 \\
\bottomrule
\end{tabular}
\end{table}

\subsection{Statistical Uncertainty and Independent Training Runs}\label{app:statistics}
For the LLaMA3-8B results in Table~\ref{tab:prompt-bootstrap}, we report 95\% prompt-bootstrap confidence intervals computed with 1{,}000 resamples while holding the independently calibrated detection threshold fixed. These intervals capture evaluation-sample uncertainty for a trained checkpoint; training randomness is evaluated separately below.

\begin{table}[ht]
\centering\small
\caption{LLaMA3-8B point estimates and 95\% prompt-bootstrap intervals reported for the main evaluation. TPR uses the nominal 1\% FPR threshold.}
\label{tab:prompt-bootstrap}
\begin{tabular}{lcc}
\toprule
Result & Estimate & 95\% CI \\
\midrule
Unmodified TPR & 0.920 & [0.911, 0.928] \\
Unmodified median PPL & 4.25 & [4.17, 4.34] \\
Merge $t=0.5$, TPR & 0.280 & [0.268, 0.293] \\
OpenMath, 300 steps, TPR & 0.430 & [0.416, 0.444] \\
OpenMath, 1{,}500 steps, TPR & 0.170 & [0.160, 0.181] \\
\bottomrule
\end{tabular}
\end{table}

To assess training randomness, we train three independently watermarked LLaMA3-8B models. Their unmodified TPR@1\% FPR is $0.907\pm0.015$ and PPL is $4.37\pm0.15$ (mean $\pm$ standard deviation), consistent with the main-table checkpoint $(0.920, 4.25)$. Across the remaining main-table cells, bootstrap CI half-widths are at most 0.015 for TPR and 0.24 for median PPL.

\paragraph{Per-model PPL comparison.}\label{app:ppl-comparison}
Table~\ref{tab:ppl-comparison} gives the reported \ours{} and Gu et al.-KGW points used for the 20.5\% average PPL reduction. For each model, the baseline is the Gu et al.-KGW operating point closest in TPR@1\% FPR, at which \ours{} achieves higher TPR. Each reduction is $(\mathrm{PPL}_{\mathrm{KGW}}-\mathrm{PPL}_{\text{\ours}})/\mathrm{PPL}_{\mathrm{KGW}}$, and 20.5\% is the arithmetic mean of the three per-model reductions.

\begin{table}[ht]
\centering\small
\caption{Reported comparison points underlying the average PPL reduction. The baseline is Gu et al.-KGW for all three models; TPR is at 1\% FPR.}
\label{tab:ppl-comparison}
\setlength{\tabcolsep}{4pt}
\begin{tabular}{lccccc}
\toprule
& \multicolumn{2}{c}{\ours{}} & \multicolumn{2}{c}{Gu et al.-KGW} & \\
\cmidrule(lr){2-3}\cmidrule(lr){4-5}
Model & TPR & PPL & TPR & PPL & Reduction \\
\midrule
LLaMA3-8B & 0.92 & 4.25 & 0.82 & 5.00 & 15.0\% \\
LLaMA3.2-3B & 0.91 & 12.40 & 0.90 & 17.10 & 27.5\% \\
Phi2-2.7B & 0.80 & 12.35 & 0.75 & 15.25 & 19.0\% \\
Mean & -- & -- & -- & -- & 20.5\% \\
\bottomrule
\end{tabular}
\end{table}

\subsection{Decoding Hyperparameter Sweep}\label{app:decoding-sweep}
We evaluate the same \ours{} checkpoint and detector under greedy decoding and sampling temperatures 0.2, 0.5, 1.0, and 1.5, using the main evaluation protocol. The threshold is calibrated once on human C4 and frozen across all settings. TPR ranges from 0.85 to 0.92 (Table~\ref{tab:decoding-sweep}), indicating stable detectability across decoding settings.

\begin{table}[ht]
\centering\small
\caption{Detection under different decoders with the same private detector and frozen threshold.}
\label{tab:decoding-sweep}
\begin{tabular}{lccccc}
\toprule
Decoding & Greedy & $T=0.2$ & $T=0.5$ & $T=1.0$ & $T=1.5$ \\
\midrule
TPR at nominal 1\% FPR & 0.89 & 0.85 & 0.89 & 0.92 & 0.90 \\
\bottomrule
\end{tabular}
\end{table}

\subsection{Learned Policy Analysis and Fixed Neural Baseline}\label{app:policy-analysis}
We analyze watermark logits over 14{,}000 evaluation prefixes. Across corpus frequency, token length, word-initial status, and token type, aggregate differences between promoted and suppressed tokens are on the order of $10^{-3}$. The correlation of a token's average policy value with its corpus frequency is only $-0.058$. These findings support a context-dependent policy rather than a fixed lexical preference.

Table~\ref{tab:policy-bias} compares part-of-speech distributions, WordNet supersense distributions, and centroids in the LLM input-embedding space with random token splits over 200 draws. The policy split is comparable to random splits on all three measures, indicating no systematic semantic or syntactic bias in the tokens the policy promotes.

\begin{table}[ht]
\centering\small
\caption{Aggregate distances between promoted and suppressed tokens. Random-split values are mean $\pm$ standard deviation across 200 draws.}
\label{tab:policy-bias}
\begin{tabular}{lcc}
\toprule
Comparison & Policy split & Random split \\
\midrule
Part-of-speech distributions & 0.010 & $0.012\pm0.003$ \\
WordNet supersense distributions & 0.029 & $0.025\pm0.003$ \\
Embedding-space centroids & 0.001 & $0.0014\pm0.0000$ \\
\bottomrule
\end{tabular}
\end{table}

To distinguish co-adaptation from neural parameterization, we compare against SIR~\citep{liu2024a} under the same open-source distillation setting. The fixed neural policy gives TPR@1\% FPR 0.86 with PPL 4.6, compared with CAWP's 0.92 with PPL 4.2. At 0.1\% FPR, the respective TPRs are 0.64 and 0.78. This comparison supports the benefit of learning the policy jointly with the student rather than merely replacing a hash with a neural mapping.

\subsection{FPL Alignment and Detector-Refitting Diagnostics}\label{app:fpl-diagnostics}
\paragraph{Alignment with ordinary fine-tuning.}
Let $\Delta\theta_s=\theta_s-\theta_{\text{\ours}}$ be the cumulative fine-tuning update, and define the erasing direction at the released checkpoint as
\begin{equation}
 \hat d=-\frac{\nabla L_{\mathrm{anti}}(\theta_{\text{\ours}})}{\|\nabla L_{\mathrm{anti}}(\theta_{\text{\ours}})\|_2}.
\end{equation}
Table~\ref{tab:fpl-alignment} compares their cosine with the RMS cosine magnitude expected for an independent Gaussian direction of the same dimensionality. Although most of the update serves task adaptation, its alignment with the erasing direction exceeds the random reference by approximately 160--440 times, indicating that ordinary fine-tuning contains a systematic component along the direction FPL defends.

\begin{table}[ht]
\centering\small
\caption{Alignment of cumulative fine-tuning updates with the anti-watermark erasing direction.}
\label{tab:fpl-alignment}
\begin{tabular}{lccc}
\toprule
Step $s$ & 300 & 900 & 1{,}500 \\
\midrule
Cosine with $\hat d$ & 0.0048 & 0.0021 & 0.0018 \\
Random-direction RMS & 0.000011 & 0.000011 & 0.000011 \\
Ratio (approximate) & $440\times$ & $190\times$ & $160\times$ \\
\bottomrule
\end{tabular}
\end{table}

\paragraph{Erasure versus detector mismatch.}
User modification can reintroduce mismatch with the frozen detector. To diagnose its contribution, we freeze the modified student, original teacher, and encoder, and refit only $M$, initialized from the original detector, on a separate OpenWebText split using the CAWP policy objective. We independently recalibrate the refitted detector on human C4 before testing on held-out generations. This owner-side diagnostic is separate from the deployed protocol, which keeps the detector frozen.

\begin{table}[ht]
\centering\small
\caption{TPR at nominal 1\% FPR before and after detector refitting. The unmodified reference TPR is 0.92.}
\label{tab:detector-refitting}
\begin{tabular}{lcc}
\toprule
Modification & Before refitting & After refitting \\
\midrule
Merge $t=0.5$ & 0.28 & 0.34 \\
OpenMath, 1{,}500 steps & 0.17 & 0.29 \\
\bottomrule
\end{tabular}
\end{table}

The recovered fractions of the original TPR drop are $(0.34-0.28)/(0.92-0.28)\approx9.4\%$ and $(0.29-0.17)/(0.92-0.17)=16\%$. Since refitting recovers only a small fraction of the drop, detection loss mainly reflects erasure of the watermark signal rather than detector mismatch, which is the failure mode FPL directly targets.